\documentclass[journal]{IEEEtran}
\usepackage{amsmath,amsfonts}
\usepackage{algorithmic}
\usepackage{array}
\usepackage{tabularx} % 自动计算列宽
\usepackage[caption=false,font=normalsize,labelfont=sf,textfont=sf]{subfig}
\usepackage{textcomp}
\usepackage{stfloats}
\usepackage{url}
\usepackage{color}
\usepackage[table]{xcolor}
\usepackage{verbatim}
\usepackage{graphicx}
\usepackage{subfig}
\usepackage[absolute,overlay]{textpos}
\def\BibTeX{{\rm B\kern-.05em{\sc i\kern-.025em b}\kern-.08em
		T\kern-.1667em\lower.7ex\hbox{E}\kern-.125emX}}
\usepackage{balance}
\usepackage{cite}
\usepackage{hyperref}
\usepackage{amsthm}  % 引入amsthm包以支持定理环境

\begin{document}
	
	%标题
	\title{DJSCC-Enabled Multi-User Semantic CSI Feedback for Hybrid Beamforming in Dual-Polarized cmWave Massive MIMO}
	
	%作者
	\author{Ziqi Han, Ziwei Wan, Hengwei Zhang, Keke Ying, Chabalala S. Chabalala, Dapeng Li, Wei Wang, and Zhen Gao
	
%	\thanks{The work was supported in part by the Natural Science Foundation of
%		China (NSFC) under Grant 62471036, Beijing Natural Science Foundation
%		under Grant L242011, Shandong Province Natural Science Foundation
%		under Grant ZR2022YQ62, and Beijing Nova Program. (\textit{Corresponding
%		author: Zhen Gao})
%	}
%	
\thanks{The work was supported in part by the Shandong Province Natural Science Foundation under Grant ZR2025QA30, in part by Natural Science Foundation of China (NSFC) under Grant 62471036 and Grant U2233216, in part by Beijing Natural Science Foundation under Grants L242011, QY24167, QY25256, QY25257. (Corresponding author: Zhen Gao and Dapeng Li.)}
\thanks{Ziqi Han and Keke Ying are with the School of Information and Electronics and the Advanced Research Institute of Multidisciplinary Science, Beijing Institute of Technology, Beijing 100081, China (e-mail: hanziqi@bit.edu.cn, ykk@bit.edu.cn).}
\thanks{Ziwei Wan and Hengwei Zhang are with the Yangtze Delta Region Academy, BIT (Jiaxing), Jiaxing 314019, China, and also with the School of Information and Electronics and the Advanced Research Institute of Multidisciplinary Sciences, Beijing Institute of Technology, Beijing 100081, China (e-mail: ziweiwan@bit.edu.cn, zhanghengwei2001@163.com).}
\thanks{Chabalala S. Chabalala is with School of Electrical and Information Engineering, University of the Witwatersrand, Johannesburg 2050, South Africa (e-mail: chabalala.chabalala@wits.ac.za).}
\thanks{Dapeng Li is with the State Key Laboratory of Environment Characteristics and Effects for Near-Space, School of Optics and Photonics, Beijing Institute of Technology, Beijing 100081, China (e-mail: dapangli@bit.edu.cn).}
\thanks{Wei Wang is with the Harbin Institute of Technology, Shenzhen 518055, China (e-mail: wang\_wei@hit.edu.cn).}
\thanks{Zhen Gao is with  Beijing Institute of Technology (BIT), Zhuhai 519088, China, also with the State Key Laboratory of CNS/ATM, Beijing 100081, China, also with the MIIT Key Laboratory of Complex-Field Intelligent Sensing, Beijing 100081, China, also with the Advanced Technology Research Institute, BIT, Jinan 250307, China, also with the Yangtze Delta Region Academy, BIT, Jiaxing 314019, China, also with Shaanxi Key Laboratory of Information Communication Security, Xi'an University of Posts Network \& Telecommunications, Xi'an, Shaanxi 710121, China (e-mail: gaozhen16@bit.edu.cn).
}
\thanks{Copyright (c) 2026 IEEE. Personal use of this material is permitted. However, permission to use this material for any other purposes must be obtained from the IEEE by sending a request to pubs-permissions@ieee.org.}

}
	%致谢
	%\thanks{Manuscript created October, 2020; This work was developed by the IEEE %Publication Technology Department. This work is distributed under the \LaTeX \ %Project Public License (LPPL) ( http://www.latex-project.org/ ) version 1.3. A %copy of the LPPL, version 1.3, is included in the base \LaTeX \ documentation of %all distributions of \LaTeX \ released 2003/12/01 or later. The opinions %expressed here are entirely that of the author. No warranty is expressed or %implied. User assumes all risk.}}

%\markboth{Journal of \LaTeX\ Class Files,~Vol.~18, No.~9, September~2020}%
%{How to Use the IEEEtran \LaTeX \ Templates}

\maketitle

\begin{abstract}	
Driven by the ultra-high throughput requirements of 6G, wireless communications are migrating to centimeter wave (cmWave) bands to overcome the limitations of current spectral resources. Massive multiple-input multiple-output (MIMO) and orthogonal frequency division multiplexing (OFDM) systems aim to achieve high spectral efficiency in cmWave regimes but are often constrained by the heavy overhead of downlink channel state information (CSI) feedback. This paper proposes a deep learning scheme based on the multi-axis multi-layer perceptron for image processing (MAXIM) architecture for joint semantic CSI feedback and hybrid beamforming in multi-user cmWave MIMO-OFDM systems, which maximizes the downlink sum rate by end-to-end optimization. Specifically, distributed encoders at multiple user equipments (UEs) perform limited CSI feedback, while the decoder at the base station (BS) jointly designs the hybrid beamforming matrices without explicit CSI reconstruction. The uplink transmission is implemented via deep joint source–channel coding (DJSCC) to enhance CSI compression efficiency and noise robustness. Furthermore, considering the high correlation between vertical and horizontal polarization channels in dual-polarized massive MIMO systems, a cross-polarization interaction module is introduced at the UEs to exploit polarization correlations for joint CSI compression. Simulation results demonstrate that the proposed method improves the downlink sum rate under various signal-to-noise ratio (SNR) conditions with a limited number of feedback symbols, validating its robustness and superiority in multi-user dual-polarized cmWave MIMO-OFDM systems.

\end{abstract}


\section{Introduction}
\subsection{Background}
\IEEEPARstart{M}{assive} multiple-input multiple-output (MIMO) has become a key technology for future wireless communication systems, particularly as network requirements evolve toward 6G standards. To meet the growing demand for ultra-high throughput and multi-gigabit-per-second (Gbps) wireless connectivity, future wireless systems are increasingly exploring the centimeter wave (cmWave) spectrum (typically 3–30 GHz) due to its relatively abundant bandwidth resources \cite{2025wangcmwave,al2019millimetre}. Compared with the millimeter wave (mmWave) spectrum (typically 30–300 GHz), which has been extensively studied in the literature \cite{li2016mmwave,wang2022mmwave,cai2020mmwave}, cmWave offers a more favorable tradeoff between capacity and coverage, making it a promising spectrum candidate for 6G wireless networks \cite{al2019millimetre}. In these high frequency regimes, massive MIMO is indispensable for providing the necessary beamforming gains to compensate for severe path loss \cite{2025CmWave_Sub-THz}, thereby significantly improving spectral efficiency and link reliability without requiring additional bandwidth or transmit power. This is particularly critical for large-scale Internet of Things (IoT) deployments, such as Industrial IoT (IIoT) and smart city infrastructures, where base stations (BSs) must support a dense cluster of heterogeneous and delay-sensitive devices under stringent resource constraints. In such scenarios, the efficiency and timeliness of channel state information (CSI) feedback directly determine the system's ability to maintain high-throughput connectivity for dense sensor networks \cite{ Lee2023IoT_MIMO}. To fully exploit the potential of massive MIMO systems, the BS requires accurate downlink CSI for precoding design. In frequency division duplexing (FDD) massive MIMO systems, the user equipment (UE) estimates the downlink CSI and then feeds it back to the BS \cite{guo2022overview}. In practical communication scenarios where a BS simultaneously serves multiple users, the aggregate feedback overhead scales linearly with both the number of antennas at the BS and the number of users. Consequently, effective compression of downlink CSI is required to enhance the efficiency of the uplink transmission. After reconstructing the multi-user downlink CSI from the compressed feedback received from the UEs, the BS applies beamforming to improve the throughput of the multi-user MIMO system.
\subsection{Related Works}
This section reviews existing works across four interconnected dimensions. First, we discuss the evolution of CSI feedback \cite{kuo2012compressive,wen2018csinet,mashhadi2020distributed,gjj2024deep,wang2025enhanced} and hybrid beamforming \cite{el2014spatially,zhang2022deep} from conventional algorithms to DL-empowered techniques. Second, we analyze the transition of transmission design paradigms from separate source-channel coding (SSCC) to deep joint source-channel coding (DJSCC) for addressing feedback channel impairments \cite{xu2022adjscc,bourtsoulatze2019djscc,zhang2025djscc}. Finally, we highlight the recent shift from separate modular designs toward jointly optimized feedback and beamforming frameworks \cite{gao2022data,guo2024deep,wang2022transformer,wei2022distributed,xue2022integrated}, providing the theoretical foundation for the end-to-end semantic communication architecture proposed in this paper.

Conventional CSI feedback schemes are realized through codebook-based methods \cite{5GNR2020study} and compressed sensing (CS)-based methods \cite{kuo2012compressive}. In codebook-based CSI feedback, the feedback overhead increases dramatically as the number of antennas grows, since the size of codebook expands correspondingly. CS-based CSI feedback requires the CSI to exhibit sparsity in a specific transform domain, which consequently leads to performance degradation in practical communication scenarios \cite{mashhadi2020distributed}. Deep learning (DL) has demonstrated remarkable potential in CSI feedback tasks. In DL-based CSI feedback, CSI is treated as an image source for compression \cite{wen2018csinet}. The authors of \cite{mashhadi2020distributed} develop DeepCMC, a convolutional neural network (CNN)-based framework for distributed CSI compression and recovery in multi-user MIMO scenarios, incorporating both quantization and entropy coding within the uplink feedback process. In \cite{gjj2024deep}, a multi-layer perceptron (MLP)-based framework named RIS-CoCSINet is proposed for multi-user CSI feedback in reconfigurable intelligent surface (RIS)-assisted systems. It decouples the CSI of neighboring UEs into shared and individual features, utilizing an additional combination neural network at the BS for feature reconstruction. The authors of \cite{wang2025enhanced} propose CoTransNet, a transformer-based framework that decomposes neighboring users' CSI into shared and UE-specific components, utilizing a cooperative decoder at the BS to exploit spatial correlations.

Most existing CSI feedback schemes adopt a SSCC architecture for uplink CSI transmission, which may suffer from the “cliff effect” in real communication scenarios \cite{xu2022adjscc}. Compared with the SSCC architecture, the DJSCC architecture leverages the powerful nonlinear representation capability of neural networks to achieve more reliable CSI reconstruction during practical uplink transmission \cite{bourtsoulatze2019djscc}. The authors of \cite{zhang2025djscc} propose a DJSCC-based multi-user CSI feedback framework utilizing a residual cross-attention transformer architecture at the BS to efficiently fuse complementary channel features from nearby users, achieving high-fidelity reconstruction with significantly reduced feedback overhead.
% \begin{comment}
	\begin{table*}[htbp]
		\centering
		\renewcommand{\arraystretch}{1.3}
		\caption{Comparison of related works with our work.}
		\normalsize
		\newcolumntype{M}[1]{>{\centering\arraybackslash}m{#1}}
		\resizebox{\textwidth}{!}{
			\begin{tabular}{|M{38pt}|M{60pt}|M{48pt}|M{30pt}|M{80pt}|M{130pt}|M{50pt}|M{100pt}|}
				\hline 
				\textbf{Ref.} & \textbf{Functions} & \textbf{Backbone} & \textbf{DJSCC} & \textbf{Uplink Channel Type} & \textbf{Beamforming Scheme} & \textbf{Uplink Feedback Time Slots} & \textbf{Optimization Objective}\\
				\hline
				\cite{mashhadi2020distributed} & CSI feedback & CNN & ${\times}$ & Perfect bit feedback & \textbackslash & $>1$ & NMSE minimization\\
				\hline
				\cite{gjj2024deep} & CSI feedback & MLP & ${\times}$ & Perfect bit feedback & \textbackslash & $>1$ & NMSE minimization\\
				\hline
				\cite{wang2025enhanced} & CSI feedback & Transformer & ${\times}$ & Perfect bit feedback & \textbackslash & $>1$ & NMSE minimization \& SGCS maximization\\
				\hline
				\cite{zhang2025djscc} & CSI feedback & Transformer & \checkmark & AWGN channel & \textbackslash & $>1$ & NMSE minimization\\

				\hline
				\cite{gao2022data} & Joint pilot transmission, CSI feedback and hybrid beamforming & MLP & ${\times}$ & AWGN channel & DL-based hybrid beamforming & $>1$ & Sum-rate maximization\\
				\hline
				\cite{guo2024deep} & Joint CSI feedback and digital beamforming & CNN & \checkmark & Perfect feedback for eigenvalues \& Realistic MIMO for eigenvectors & DL-based fully-digital beamforming & $>1$ & Sum-rate maximization\\
				\hline
				\cite{wang2022transformer} & Joint CSI feedback and hybrid beamforming & Transformer & ${\times}$ & Perfect bit feedback & DL-based hybrid beamforming & $>1$ & Sum-rate maximization\\
				\hline
				\cite{wei2022distributed} & Joint CSI feedback and hybrid beamforming & CNN & ${\times}$
				& AWGN channel & DL-based hybrid beamforming & $>1$ & Sum-rate maximization\\
				\hline
				\cite{xue2022integrated} & Joint CSI feedback and hybrid beamforming & CNN & ${\times}$ & AWGN channel & Hybrid: DL-based analog beamforming \& ZF-based digital beamforming & $>1$ & Sum-rate
				maximization\\
				\hline
				our work & Joint CSI feedback and hybrid beamforming & MAXIM & \checkmark & Realistic MIMO-OFDM channel & DL-based hybrid beamforming & 1 & Sum-rate maximization\\
				\hline
		\end{tabular}}
		\label{survey_tab}
	\end{table*}

Accurate CSI at the BS is essential for downlink precoding design and throughput enhancement in massive MIMO systems. To reduce the hardware cost and power consumption of fully digital architectures, hybrid beamforming has been widely adopted in massive MIMO systems \cite{So2017hbf}. This advantage is particularly appealing for resource-constrained IoT communication scenarios in the cmWave band, where high beamforming gain is desired under stringent implementation constraints \cite{kunwar2022IoT_HBF}. However, the constant-modulus constraint of analog phase shifters renders the hybrid beamforming design a challenging non-convex optimization problem \cite{el2014spatially}. Compared with conventional iterative beamforming algorithms, DL-based beamforming schemes can approximate the optimal performance with much lower computational complexity \cite{zhang2022deep}.

Despite the advantages of DL-based beamforming, its performance remains fundamentally limited by the accuracy of CSI feedback. To further mitigate this limitation, recent research has shifted from individual module optimization toward the joint design of CSI feedback and beamforming. Compared with separately optimizing these two modules, end-to-end joint optimization directly maps the received feedback symbols to the beamforming matrices without explicit CSI reconstruction. Under this paradigm, the feedback is redefined as semantic CSI, consisting of task-relevant features extracted from raw channel matrices that characterize the spatial properties requisite for beamforming.\footnote{Formally, the task-relevant feature $\mathbf{z}$ is characterized via the Information Bottleneck principle \cite{tishby2000information_bottle} as the compressed representation of the channel $\mathbf{H}$, that minimizes the objective $\mathcal{L} = \min[I(\mathbf{H}; \mathbf{z}) - \beta I(\mathbf{z}; \mathbf{W}_{\text{opt}})]$, where $I(\cdot;\cdot)$ denotes mutual information and $\beta$ is the Lagrange multiplier controlling the trade-off between compression and task performance. And $\mathbf{W}_{\text{opt}}$ denotes the optimal hybrid beamforming matrix required for the downstream task.} Accordingly, semantic compression aims to extract and transmit only this task-relevant semantic information at the transmitter \cite{Deniz2023semantic}. Unlike conventional CSI, semantic CSI filters out redundant channel components to prioritize information that maximizes the communication objective, such as the downlink sum rate \cite{zyf2024isac,lsc2023csab}. This approach overcomes the performance bottleneck caused by objective mismatch and achieves superior system performance under limited feedback overhead \cite{gao2022data}. Existing works have explored various DL architectures to optimize joint CSI feedback and beamforming. Initial efforts, such as the end-to-end MLP-based network proposed in \cite{gao2022data}, jointly model pilot transmission, feedback, and beamforming. However, simple MLP structures struggle to capture the complex spatial correlations of high-dimensional MIMO channels. To address this, CNN-based architectures are introduced to exploit spatial features more effectively. For instance, DNet in \cite{wei2022distributed} employs distributed neural networks to perform multi-user CSI compression and learns hybrid beamforming matrices from the reconstructed CSI under limited feedback conditions. Similarly, the CsiBFNet framework in \cite{xue2022integrated} utilizes CNNs to optimize CSI feedback and analog beamforming. Nevertheless, it still relies on the conventional zero-forcing (ZF) algorithm for digital precoding, which limits the potential of end-to-end optimization. To overcome the local receptive field limitations of CNNs, the authors of \cite{wang2022transformer} introduce a transformer-based scheme that leverages the self-attention mechanism to capture global spatial dependencies across large antenna arrays. Despite the enhanced representation learning capabilities offered by these architectures, existing frameworks typically assume idealized uplink feedback links, thereby neglecting the practical channel impairments inherent in transmission environments. To address this limitation, the authors of \cite{guo2024deep} recently propose a DJSCC-based joint CSI feedback and multi-user precoding scheme, which specifically enhances the robustness of CSI feedback against channel noise and link degradation.

Moreover, \cite{gao2022data,wei2022distributed,xue2022integrated} assume an additive white Gaussian noise (AWGN) uplink channel, while \cite{wang2022transformer} considers only quantization loss, thereby neglecting the impact of the practical uplink propagation environment. In practical cmWave massive MIMO and orthogonal frequency division multiplexing (OFDM) systems, antenna arrays typically employ dual-polarized antennas, where the inherent correlation between vertical and horizontal polarizations can be leveraged for joint CSI compression \cite{han2025transformer}. In addition, most existing works assume that uplink transmission of multiple users is realized through time-division multiple access (TDMA), leading to significant latency and inefficient utilization of time–frequency resources in practical massive MIMO systems. For
clarity, the comparison of the related works is summarized in Table \ref{survey_tab}.

Moreover, the inherent capacity of our proposed framework to mitigate channel aging is primarily attributed to the exploitation of spatial semantic features, represented by the angular power spectrum, which characterizes the intrinsic structural properties of the wireless propagation environment. Unlike instantaneous complex channel coefficients that exhibit rapid fluctuations due to small-scale fading, these statistical spatial features typically exhibit relative stationarity over a significantly longer time scale. While the specific phases and amplitudes of multipath components are highly sensitive to delay, the underlying spatial structure, such as the directions of departure, evolves slowly. By prioritizing these stable spatial semantic features, our framework captures the essential directional information required for beamforming, thereby maintaining high beamforming gain even when the feedback is subject to the latencies commonly encountered in contention-based IoT access or heterogeneous processing environments.

\subsection{Our Contributions}
Motivated by the limitations discussed above, we propose a DJSCC-based joint CSI feedback and hybrid beamforming network for multi-user dual-polarized cmWave MIMO-OFDM systems. The main contributions are summarized as follows:
\begin{itemize}
	\item We propose an end-to-end multi-axis MLP for image processing (MAXIM)\cite{tu2022maxim}-based network for the joint design of multi-user semantic CSI feedback and hybrid beamforming. By avoiding explicit CSI reconstruction and directly mapping received uplink signals to beamforming matrices, the proposed framework effectively mitigates error propagation and focuses on extracting task-oriented features to maximize hybrid beamforming gain.
	\item We establish a multi-user DJSCC framework that supports simultaneous non-orthogonal uplink transmission. By integrating minimum mean square error (MMSE)-based multiple user detection (MUD) with the proposed DL architecture, our design effectively mitigates inter-user interference and channel impairments within the multi-user DJSCC framework, ensuring reliable recovery of channel semantics for beamforming in realistic massive MIMO scenarios.
	\item We develop a cross-polarization interaction (CPI) module that employs a co-attention mechanism to exploit the bidirectional correlation between dual-polarized CSI. This facilitates efficient cross-polarization feature interaction, enabling the encoder to capture synergistic features and improve joint compression performance.
	\item While conventional methods aim for accurate reconstruction of CSI images, raw pixel values are highly sensitive to time variations; even slight channel changes can lead to degradation in reconstruction quality (e.g.,	NMSE). In contrast, our scheme extracts task-oriented semantic features that capture the essential spatial structure required for beamforming, which is typically less sensitive than the raw CSI representation to time variations. Moreover, by avoding the explicit CSI reconstruction stage, the processing latency at the BS is further reduced.
	
\end{itemize}   
\textit{Notations:} Boldface lower and upper-case symbols denote column vectors and matrices, respectively. $\left[ \mathbf{A} \right]_{i,j}$ denotes the $i$-th row and $j$-th column
element of $\mathbf{A}$. Superscripts ${\left(  \cdot  \right)^{\rm{-1}}}$, ${\left(  \cdot  \right)^{\rm{T}}}$ and ${\left(  \cdot  \right)^{\rm{H}}}$ denote the inversion, transpose and conjugate transpose operators, respectively. $\left| \cdot \right|$ denotes the modulus operation and $\Vert \cdot \Vert_\mathrm{F}$ represents the Frobenius norm operation. ${\mathbf{I}_n}$ denotes the ${n\times n}$ identity matrix. The expectation is denoted by $\mathbb{E}\left(  \cdot  \right)$.

% \vspace{-20pt}
% 设置出版标识
%\IEEEpubid{This work is licensed under a Creative Commons Attribution 4.0 License.}

% 调整出版标识位置
%\IEEEpubidadjcol

\section{System Model}
We consider an FDD multi-user cmWave MIMO-OFDM system with ${N_c}$ subcarriers, which represents a multi-user IoT communication scenario. A dual-polarized antenna consists of a single physical unit with two ports, corresponding to horizontal and vertical polarizations, respectively. The BS is equipped with ${N_t}$ antenna ports (arranged as ${\frac{N_t}{2}}$ dual-polarized antennas) and ${N_\mathrm{RF}}$ RF chains, serving ${K}$ simultaneous IoT UEs. Each user is equipped with ${N_r}$ antenna ports (arranged as ${\frac{N_r}{2}}$ dual-polarized antennas). In this framework, we jointly investigate the uplink CSI feedback over the physical uplink channel and the subsequent downlink hybrid beamforming design. The downlink CSI can be denoted as ${\mathbf{H}_\mathrm{d}\in \mathbb{C}^{K\times N_c\times N_r\times N_t}}$, and the corresponding uplink CSI can be expressed as ${\mathbf{H}_\mathrm{u}\in \mathbb{C}^{K\times N_c\times N_t\times N_r}}$. It is assumed that perfect uplink CSI and downlink CSI can be acquired after the channel estimation at the BS and the UEs, respectively. 

Under the assumption of a uniform linear array configuration with half-wavelength antenna spacing at both the BS and UEs, the downlink channel matrix ${\mathbf{H}_{\mathrm{d}}\left[ k,n \right]\in \mathbb{C}^{N_r\times N_t}}$ for the $k$-th UE on the $n$-th subcarrier can be expressed as
\begin{equation}
	\mathbf{H}_{\mathrm{d}}\left[ k,n \right] = \sqrt{\frac{N_tN_r}{L_k}} \sum_{l=1}^{L_k} \beta_{l,k} e^{-j 2\pi \frac{n}{N_c} \tau_{l,k} f_s} \mathbf{d}_r \left( \psi_{l,k} \right)\mathbf{d}^{\mathrm{H}}_{t}(\phi_{l,k}),
\end{equation}
where $L_k$ denotes the number of downlink multipath components for the $k$-th user; $\tau_{l,k}$ and $\beta_{l,k}$ represent the delay and propagation gain of the $l$-th multipath component, respectively; and $f_s$ is the sampling frequency. $\psi_{l,k}$ and $\phi_{l,k}$ denote the angle of arrival and the angle of departure for the $l$-th path, respectively. $\mathbf{d}_{r}(\psi_{l,k}) = \left[ 1,\, e^{-j\pi\sin\psi_{l,k}},\, \dots,\, e^{-j\pi(N_r-1)\sin\psi_{l,k}} \right]^{\mathrm{T}}$ denotes the steering vector at the receiver for the $l$-th path, and $\mathbf{d}_{t}(\phi_{l,k}) = \left[ 1,\, e^{-j\pi\sin\phi_{l,k}},\, \dots,\, e^{-j\pi(N_t-1)\sin\phi_{l,k}} \right]^{\mathrm{T}}$ represents the steering vector at the transmitter for the $l$-th path.

The transmit signal $\mathbf{t}[n]\in\mathbb{C}^{N_t\times 1}$ at the BS on the $n$-th subcarrier can be expressed as
\begin{equation}
	\label{Eq:transmit_signal}
	\mathbf{t}[n] = \sum_{k=1}^{K} \mathbf{F}_{\mathrm{RF}} \mathbf{f}_{\mathrm{BB}}[k, n] x[k, n] = \mathbf{F}_{\mathrm{RF}} \mathbf{F}_{\mathrm{BB}}[n] \mathbf{x}[n],
\end{equation}
where ${\mathbf{F }_{\mathrm{RF}}\in \mathbb{C}^{N_t\times N_{\mathrm{RF}}}}$ is the analog beamforming matrix. The overall digital beamformer across all subcarriers is denoted by ${\mathbf{F}_{\mathrm{BB}}\in \mathbb{C}^{N_c\times N_{\mathrm{RF}}\times K}}$. For the $n$-th subcarrier, the digital beamforming matrix ${\mathbf{F}_{\mathrm{BB}}\left[n\right] \in \mathbb{C}^{N_{\mathrm{RF}}\times K}}$ is formed by concatenating the digital beamforming vectors of all $K$ users as ${\mathbf{F}_{\mathrm{BB}}\left[ n \right] =\left[ \mathbf{f}_{\mathrm{BB}}\left[ 1,n \right] ,\dots,\mathbf{f}_{\mathrm{BB}}\left[ K,n \right] \right]}$. Here, ${\mathbf{f}_{\mathrm{BB}}\left[ k,n \right] \in \mathbb{C}^{N_{\mathrm{RF}}\times 1}}$ represents the specific digital beamforming vector for the $k$-th user on the $n$-th subcarrier. ${x\left[ k,n \right] }$ is the transmitted data associated with the ${k}$-th user delivered on the ${n}$-th subcarrier, and ${\mathbf{x}[n] = \left[ x[1,n], x[2,n], \dots, x[K,n] \right]^{\mathrm{T}} \in \mathbb{C}^{K \times 1}}$, where $\mathbb{E}\left( \mathbf{x}[n] \mathbf{x}^{\mathrm{H}}[n] \right) = \mathbf{I}_K$ for ${n=1,2,\dots,N_c}$. In the downlink transmission phase, the received signal of the $k$-th user on the $n$-th subcarrier, denoted by ${\mathbf{r}\left[ k,n \right]\in \mathbb{C}^{N_{r}\times 1}}$, is given by
\begin{equation}
	\label{Eq:receive_signal}
	\begin{aligned}
		\mathbf{r}\left[ k,n \right] &= \mathbf{H}_{\mathrm{d}}\left[ k,n \right] \mathbf{F}_{\mathrm{RF}} \mathbf{f}_{\mathrm{BB}}\left[ k,n \right] {x}\left[ k,n \right] \\
		&+ \sum_{k' \neq k} \mathbf{H}_{\mathrm{d}}\left[ k,n \right] \mathbf{F}_{\mathrm{RF}} \mathbf{f}_{\mathrm{BB}}\left[ k',n \right] {x}\left[ k',n \right] + \mathbf{n}\left[ k,n \right],
	\end{aligned}
\end{equation}
where ${\mathbf{H}_{\mathrm{d}}\left[ k,n \right] \in \mathbb{C}^{N_r\times N_t}}$ is the downlink CSI corresponding to the ${k}$-th user on the ${n}$-th subcarrier; And ${\mathbf{n}\left[ k,n \right] \sim \mathcal{C}\mathcal{N}\left( \mathbf{0},\sigma _{d}^{2}\mathbf{I}_{N_r} \right)}$ denotes the AWGN vector. 

\begin{figure*}[t]
	\centering
	\includegraphics[width=1.009\textwidth]{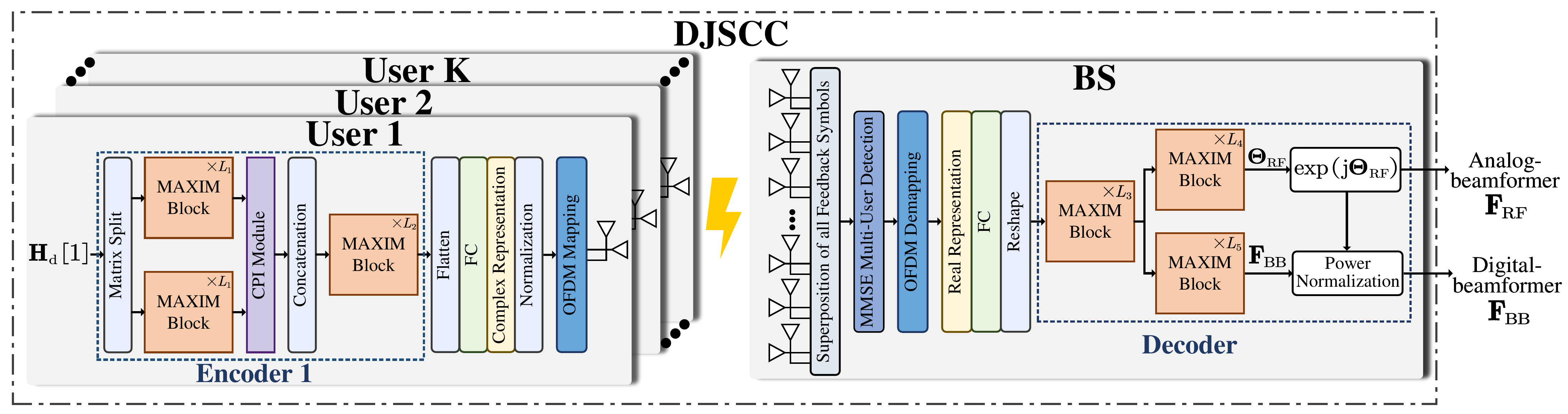}
	%\vspace{-15pt}
	\caption{Proposed DJSCC-based multi-user joint CSI feedback and hybrid beamforming network.}
	\vspace{-15pt}  % 减少图片下方的间距
	\label{fig:overall framework}
\end{figure*}

\section{Proposed Multi-user Joint CSI Feedback and Hybrid Beamforming Scheme}
In this section, we begin with the problem formulation for the proposed MAXIM-driven joint DJSCC-CSI feedback and hybrid beamforming network for multi-user dual-polarized cmWave MIMO-OFDM systems. Subsequently, we introduce the overall network architecture, and then provide the details of the CPI module which exploits cross-polarization correlations for joint dual-polarized CSI compression.

% \vspace{-10pt}
\subsection{Problem Formulation of the Proposed Scheme}
During the uplink CSI feedback phase, the DJSCC encoder at the ${k}$-th UE maps its downlink CSI ${\mathbf{H}_\mathrm{d}[k]\in \mathbb{C}^{N_c\times N_r\times N_t}}$ into a compressed semantic feature vector ${\mathbf{s}\left[ k \right] \in \mathbb{C}^m}$, which is expressed as
\begin{equation}
	\label{Eq:DISCCencoder}
	\mathbf{s}\left[ k \right] =f_{\mathrm{EN}}\left( \mathbf{H}_\mathrm{d}\left[ k \right] \right),\ k=1,\dots,K,
\end{equation}
where ${f_{\mathrm{EN}}\left( \cdot \right)}$ denotes the DJSCC encoder with trainable parameters ${\mathbf{\Psi}_\mathrm{EN}}$. The encoded vector of the ${k}$-th user ${\mathbf{s}\left[ k \right] =\left[ s\left[ k,1 \right] ,s\left[ k,2 \right] ,\dots,s\left[ k,m \right] \right] ^\mathrm{T}}$ is mapped to $m$ orthogonal subcarriers and transmitted to the BS by the ${k}$-th user, and ${s\left[ k,n \right] }$ is the symbol associated with the ${k}$-th user on the $n$-th subcarrier. Due to the parallel transmission of multi-user uplink signals, the received signal ${\mathbf{y}\left[ n \right] \in \mathbb{C}^{N_t\times 1}}$ at the BS on the ${n}$-th subcarrier can be expressed as
\begin{equation}
	\label{Eq:received signal}
	\mathbf{y}\left[ n \right] =\sum_{k=1}^K{\mathbf{H}_\mathrm{u}\left[ k,n \right] \mathbf{v}s\left[ k,n \right]}+\mathbf{z}\left[ n \right] ,\,\,n=1,\dots,m,
\end{equation} %%%%%%%%%%
where ${\mathbf{H}_\mathrm{u}\left[ k,n \right] \in \mathbb{C}^{N_t\times N_r}}$ represents the uplink channel matrix corresponding to the ${k}$-th user on the ${n}$-th subcarrier. The term ${\mathbf{v}\in \mathbb{C}^{N_r\times 1}}$ is the transmit precoding vector at the UE. For simplicity, $\mathbf{v}$ is assumed to be an all-ones vector, thereby ensuring that an identical symbol is transmitted across all $N_r$ antenna ports. ${\mathbf{z}\left[ n \right] \sim \mathcal{C}\mathcal{N}\left( \mathbf{0},\sigma _{u}^{2}\mathbf{I}_{N_t} \right) }$ denotes the AWGN vector at the BS on the ${n}$-th subcarrier. The effective uplink channel vector for the ${k}$-th user on the ${n}$-th subcarrier is defined as ${\mathbf{h}_\mathrm{eff}\left[ k,n \right] =\mathbf{H}_\mathrm{u}\left[ k,n \right] \mathbf{v}\in \mathbb{C}^{N_t\times 1}}$. By concatenating the effective channels of all $K$ users, we define the composite uplink effective channel matrix as ${\mathbf{H}_{\mathrm{eff}}[n] = \left[ \mathbf{h}_{\mathrm{eff}}[1,n], \dots, \mathbf{h}_{\mathrm{eff}}[K,n] \right] \in \mathbb{C}^{N_t \times K}}$. Furthermore, let ${\mathbf{s}\left[ n \right] =\left[ s\left[ 1,n \right] ,\dots,s\left[ K,n \right] \right] ^{\mathrm{T}}\in \mathbb{C}^K}$ denotes the stacked signal vector on the ${n}$-th subcarrier. Therefore, the received signal in Eq. (\ref{Eq:received signal}) can be rewritten as
\begin{equation}
	\label{Eq:received signal rewrite}
	\begin{aligned}
		\mathbf{y}\left[ n \right] 
		&= \sum_{k=1}^K{\mathbf{h}_{\mathrm{eff}}\left[ k,n \right] s\left[ k,n \right]}+\mathbf{z}\left[ n \right]\\
		&= \mathbf{H}_{\mathrm{eff}}\left[ n \right] \mathbf{s}\left[ n \right] +\mathbf{z}\left[ n \right].
	\end{aligned} %%%%%%%%
\end{equation}

To mitigate multi-user interference, a linear MMSE detector is applied to the received signal on each subcarrier, which is given by
\begin{equation}
	\label{Eq:WMMSE}
	\mathbf{W}_{\mathrm{MMSE}}\left[ n \right] =\left( \mathbf{H}_{\mathrm{eff}}^{\mathrm{H}}\left[ n \right] \mathbf{H}_{\mathrm{eff}}\left[ n \right] +\sigma _{u}^{2}\mathbf{I}_{K} \right) ^{-1}\mathbf{H}_{\mathrm{eff}}^{\mathrm{H}}\left[ n \right] ,
\end{equation}
where ${\mathbf{W}_{\mathrm{MMSE}}\left[ n \right] \in \mathbb{C}^{K\times N_t}}$ represents the MMSE-based detection matrix on the ${n}$-th subcarrier. Therefore, the estimated symbol vector ${\mathbf{\hat{s}}\left[ n \right]\in \mathbb{C}^K}$ can be recovered by applying the detection matrix to the received signal ${\mathbf{y}\left[ n \right]}$, which can be expressed as
\begin{equation}
	\label{Eq:MUD}
	\mathbf{\hat{s}}\left[ n \right] =\mathbf{W}_{\mathrm{MMSE}}\left[ n \right] \mathbf{y}\left[ n \right].
\end{equation}

At the BS, the recovered semantic feature vectors from all users are aggregated to perform joint hybrid beamforming design, which can be expressed as
\begin{equation}
	\label{Eq:HBFdesign}
	\left\{\mathbf{\Theta }_{\mathrm{RF}},\mathbf{F}_{\mathrm{BB}}\right\}=f_{\mathrm{HBF}}\left( \left\{ \mathbf{\hat{s}}\left[ n \right] \right\} _{n=1}^{m} \right) ,
\end{equation}
\begin{equation}
	\label{Eq:analog BF}
	\mathbf{F}_{\mathrm{RF}}=\cos(\mathbf{\Theta }_{\mathrm{RF}}) + j \cdot \sin(\mathbf{\Theta }_{\mathrm{RF}}),
\end{equation}
where ${f_{\mathrm{HBF}}\left( \cdot \right)}$ denotes the DJSCC decoder with trainable parameters ${\mathbf{\Psi}_\mathrm{HBF}}$. ${\mathbf{\Theta }_{\mathrm{RF}}\in \mathbb{R}^{N_t\times N_{\mathrm{RF}}}}$ represents the phase values of the analog beamformer. The analog beamforming matrix ${\mathbf{F }_{\mathrm{RF}}}$ is then constructed as Eq. (\ref{Eq:analog BF}). 

According to Eq. (\ref{Eq:receive_signal}), the achievable rate of the ${k}$-th user on the ${n}$-th subcarrier can be expressed as
% \begin{equation}
	%	\label{Eq:rate}
	%	R_{k,n} = \log_2\left( 1 + \frac{\left| \mathbf{H}_{\mathrm{d}}\left[ k,n \right] \mathbf{F}_{\mathrm{RF}} \mathbf{f}_{\mathrm{BB}}\left[ k,n \right] \right|^2}{\sum_{k' \neq k} \left| \mathbf{H}_{\mathrm{d}}\left[ k,n \right] \mathbf{F}_{\mathrm{RF}} \mathbf{f}_{\mathrm{BB}}\left[ k',n \right] \right|^2 + \sigma_{d}^{2}} \right).
	%\end{equation}
	\begin{equation}
		\label{Eq:rate}
		R_{k,n}=\log _2\,\det \left( \mathbf{I}_{N_r}+\mathbf{\Phi} _{k,n}^{-1}\mathbf{H}_{\mathrm{d}}\left[ k,n \right] \mathbf{w}_{k,n}\mathbf{w}_{k,n}^{\mathrm{H}}\mathbf{H}_{\mathrm{d}}^{\mathrm{H}}\left[ k,n \right] \right),
	\end{equation}
	where ${\mathbf{w}_{k,n}=\mathbf{F}_{\mathrm{RF}} \mathbf{f}_{\mathrm{BB}}\left[ k,n \right]\in\mathbb{C}^{N_t\times1}}$ is the effective precoding vector associated with the $k$-th user on the $n$-th subcarrier; ${\mathbf{\Phi} _{k,n}=\sigma _{d}^{2}\mathbf{I}_{N_r}+\sum_{k'\neq k}\mathbf{H}_{\mathrm{d}}\left[ k,n \right] \mathbf{w}_{k',n}\mathbf{w}_{k',n}^{\mathrm{H}}\mathbf{H}_{\mathrm{d}}^{\mathrm{H}}\left[ k,n \right]}$ is the interference-plus-noise covariance matrix for the $k$-th user on the $n$-th subcarrier.
	
	Then, the sum rate of the system is calculated by
	\begin{equation}
		\label{Eq:sum_rate}
		R=\frac{1}{N_c}\sum_{k=1}^K{\sum_{n=1}^{N_c}{R_{k,n}}}.
	\end{equation}
	
	Accordingly, the optimization problem of the hybrid beamforming in the multi-user massive MIMO system is formulated as
	
	\begin{equation}
		\label{Eq:max_sum_rate}
		\left\{ \mathbf{F}_{\mathrm{RF}}, \mathbf{F}_{\mathrm{BB}} \right\} = \mathop{\arg\max}\limits_{\mathbf{\Psi}_\mathrm{EN}, \mathbf{\Psi}_\mathrm{HBF}} R,
	\end{equation}
	\begin{equation}
		\label{Eq:RF_limit}
		\mathrm{s.t.} \, \left| \left[ \mathbf{F}_{\mathrm{RF}} \right]_{i,j} \right| = \frac{1}{\sqrt{N_t}}, \quad \forall i,j,
	\end{equation}
	\begin{equation}
		\label{Eq:RFBB_limit}
		\left\| \mathbf{F}_{\mathrm{RF}} \mathbf{F}_{\mathrm{BB}} \left[ n \right] \right\|_{\mathrm{F}}^2 \le K.
	\end{equation}

\subsection{DJSCC-based Multi-user Joint CSI Feedback and Hybrid Beamforming Network Architecture}
As illustrated in Fig. \ref{fig:overall framework}, in the multi-user scenario, each UE compresses its downlink CSI via a DJSCC encoder. The encoded semantic features are then mapped onto OFDM subcarriers and transmitted simultaneously over the MIMO uplink channel. At the BS, the superimposed signals from all $K$ UEs are processed by an MMSE-based MUD to suppress inter-user interference and obtain recovered symbols. These recovered symbols, which remain corrupted by uplink channel impairments and residual multi-user interference, are then demapped and fed into the DJSCC decoder. Subsequently, the DJSCC decoder jointly processes the multi-user semantic features to design the hybrid beamforming matrices for the subsequent downlink transmission. For practical implementation, a weight-sharing strategy is adopted, where all UEs employ an identical DJSCC encoder architecture and parameter set, thereby ensuring scalability and reducing training complexity.

\begin{figure*}[t]
	\centering
	\includegraphics[width=1\textwidth]{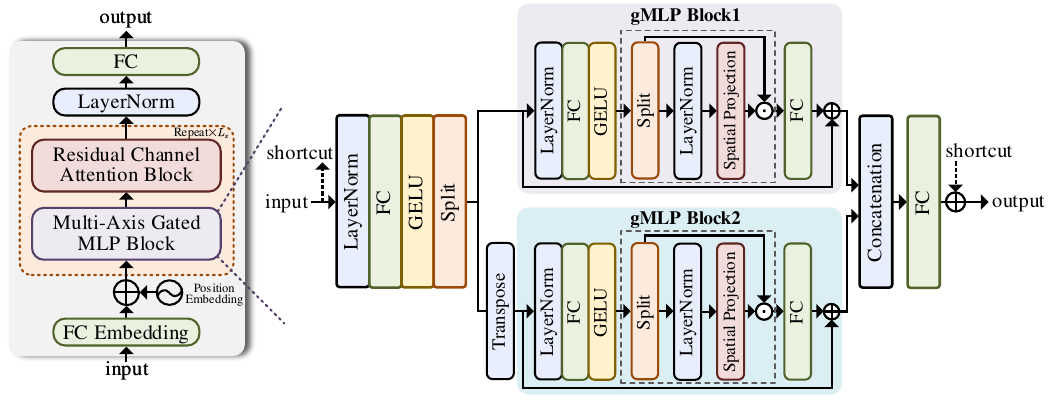}
	%\vspace{-15pt}
	\caption{Structure of the MAXIM block and the MAB.}
	\vspace{-15pt}  % 减少图片下方的间距
	\label{fig:MAXIM Block}
\end{figure*}

In massive MIMO systems, the antenna array and subcarriers share the same physical scattering environment, leading to highly correlated multipath components, such as angles of arrival and delays, across all antenna elements. Given the significant structured correlations of CSI across the frequency and antenna domains, CNN struggles to capture long-range dependencies across subcarriers or antenna ports due to its localized receptive fields. While the self-attention-based transformer has the capability to model long-range dependencies, its computational complexity scales quadratically with the input size, which is prohibitively high for real-time CSI feedback. To address this, we adopt the MAXIM network, which utilizes the multi-axis gated MLP block (MAB) to perform feature interactions across multiple orthogonal token axes, thereby effectively capturing global CSI correlations across the frequency and antenna domains with reduced computational overhead\cite{tu2022maxim}. Consequently, MAXIM is employed as the backbone for both semantic feature extraction and hybrid beamforming design in our multi-user scenario. Fig. \ref{fig:MAXIM Block} depicts the internal structure of the MAXIM block, which integrates two core modules: an MAB and a residual channel attention block (RCAB) \cite{wang2017residual}. To enhance the representational capacity of the network, these two components are cascaded for $L_x$ successive stages within each MAXIM block. The expected input to each MAXIM block is a real-valued tensor with dimensions ${\left( n_1,n_2,d_{model} \right) }$, where ${n_1}$ and ${n_2}$ represent the spatial dimensions, and ${d_{model}}$ is the embedding dimension. Specifically, the MAB is built upon the gated MLP (gMLP) mechanism. Taking gMLP block1 in Fig. \ref{fig:MAXIM Block} as an example, the input feature $\mathbf{R}_{\mathrm{in}} \in \mathbb{R}^{n_1 \times n_2 \times d_{model}}$ is processed through the following operations
\begin{equation}
	\label{Eq:gmlp1}
	\mathbf{R}_{\mathrm{1}} = \sigma\left(\mathrm{LayerNorm}(\mathbf{R}_{\mathrm{in}}) \mathbf{W}_{1} + \mathbf{b}_{1}\right),
\end{equation}
where $\mathbf{W}_1\in\mathbb{R}^{d_{model}\times 2d_{model}}$ and $\mathbf{b}_1\in\mathbb{R}^{2d_{model}}$ are the weight matrix and bias vector of the first linear layer, respectively. $\mathrm{LayerNorm}(\cdot)$ denotes layer normalization operation, and $\sigma(\cdot)$ is the Gaussian error linear unit (GELU) activation function. The primary characteristic of the gMLP block is the spatial gating unit (SGU), which is designed to enable cross-token spatial interactions. Specifically, the intermediate representation $\mathbf{R}_1\in\mathbb{R}^{n_1\times n_2\times 2d_{model}}$ is partitioned along the channel dimension into two components, $\mathbf{U}$ and $\mathbf{V}\in\mathbb{R}^{n_1\times n_2\times d_{model}}$. The gating operation is then defined as
\begin{equation}
	\label{Eq:sgu}
	\mathrm{SGU}(\mathbf{R}_{\mathrm{1}}) = \mathbf{U} \odot \left( \mathbf{W}_2 \mathrm{LayerNorm}(\mathbf{V}) + \mathbf{b}_2 \right),
\end{equation}
where $\odot$ denotes the element-wise multiplication. In this gating mechanism, $\mathbf{V}$ is processed by a linear spatial projection $\mathbf{W}_2\in\mathbb{R}^{n_2\times n_2}$ along with a bias term $\mathbf{b}_2 \in \mathbb{R}^{n_2}$ to generate a gating signal that modulates $\mathbf{U}$. This operation effectively captures long-range spatial correlations by enabling cross-token interactions without the need for a self-attention mechanism. Finally, the output of the SGU is projected back to the original input dimension and integrated with the input via a residual connection to obtain the final output $\mathbf{R}_{\mathrm{out}}\in\mathbb{R}^{n_1\times n_2\times d_{model}}$
\begin{equation}
	\label{Eq:gmlp2}
	\mathbf{R}_{\mathrm{out}} = \mathbf{R}_{\mathrm{in}} + \left(\mathrm{SGU}(\mathbf{R}_{\mathrm{1}}) \mathbf{W}_{3} + \mathbf{b}_{3}\right),
\end{equation}
where $\mathbf{W}_3\in\mathbb{R}^{d_{model}\times d_{model}}$ and $\mathbf{b}_3\in\mathbb{R}^{d_{model}}$ are the weight matrix and bias vector of the last linear layer.

For the ${k}$-th user, the downlink CSI is denoted by a complex-valued tensor ${\mathbf{H}_{\mathrm{d}}\left[ k \right] \in \mathbb{C}^{N_c\times N_r\times N_t}}$. To satisfy the real-valued input requirement of the neural network and the input shape requirement of the MAXIM block, ${\mathbf{H}_{\mathrm{d}}\left[ k \right]}$ is transformed into a real-valued tensor ${\tilde{\mathbf{H}}_{\mathrm{d}}[k] \in \mathbb{R}^{N_c \times N_t \times N_r \times 2}}$. We assume a dual-polarized configuration with $N_r=2$. Furthermore, considering the dual-polarization property, $\tilde{\mathbf{H}}_{\mathrm{d}}[k]$ can be partitioned along the receive antenna polarization dimension into two sub-tensors, ${\mathbf{H}_{\mathrm{d}}^{\mathrm{h}}\left[ k \right]}$ and ${\mathbf{H}_{\mathrm{d}}^{\mathrm{v}}\left[ k \right] \in \mathbb{R}^{N_c\times N_t\times 2}}$, representing the CSI corresponding to the horizontally and vertically polarized receive ports, respectively. Subsequently, a fully-connected (FC) embedding layer projects ${\mathbf{H}_{\mathrm{d}}^{\mathrm{h}}\left[ k \right]}$ and ${\mathbf{H}_{\mathrm{d}}^{\mathrm{v}}\left[ k \right]}$ to ${N_c\times N_t\times d_{model}}$, where ${N_c}$ and ${N_t}$ serve as the token axes, and ${d_{model}}$ represents the embedding dimension. Then, positional encoding is applied to inject information about the positions of the tokens in the sequence. Subsequently, ${\mathbf{H}_{\mathrm{d}}^{\mathrm{h}}\left[ k \right]}$ and ${\mathbf{H}_{\mathrm{d}}^{\mathrm{v}}\left[ k \right]}$ are processed separately through ${L_1}$ MAXIM blocks to extract semantic features. 
Although the horizontal and vertical ports of a dual-polarized antenna are theoretically orthogonal, ${\mathbf{H}_{\mathrm{d}}^{\mathrm{h}}\left[ k \right]}$ and ${\mathbf{H}_{\mathrm{d}}^{\mathrm{v}}\left[ k \right]}$ exhibit significant statistical correlation in practical propagation environments.
This correlation stems from factors such as mutual coupling, shared multipath components, and polarization rotation. 
As both channel components represent projections of the same spatial paths onto different polarization directions, they exhibit inherent consistency and complementarity across the spatial, frequency, and amplitude domains. Leveraging these dependencies, joint modeling and compression of ${\mathbf{H}_{\mathrm{d}}^{\mathrm{h}}\left[ k \right]}$ and ${\mathbf{H}_{\mathrm{d}}^{\mathrm{v}}\left[ k \right]}$ can more effectively exploit cross-polarization correlations, thereby facilitating the extraction of relevant CSI features.

Subsequently, the two outputs from the MAXIM blocks are fed into the CPI module to leverage the cross-polarization CSI correlation for joint compression at the UE. The output of CPI module is then concatenated along the receive antenna polarization dimension, yielding ${\mathbf{X}\left[ k \right] \in \mathbb{R}^{N_c\times N_r\times N_t\times 2}}$, which is further processed through ${L_2}$ MAXIM blocks for feature extraction. The output is flattened and compressed via an FC layer into a real-valued representation ${\mathbf{s}\left[ k \right] \in \mathbb{R}^{2m}}$. To improve the uplink transmission reliability, DJSCC is employed for CSI feedback. Specifically, ${\mathbf{s}\left[ k \right]}$ is mapped to a complex-valued vector ${\mathbf{s}_{c}\left[ k \right] \in \mathbb{C}^m}$, power-normalized, and mapped onto subcarriers for orthogonal transmission over the uplink MIMO-OFDM channel. Assuming simultaneous feedback from all users, the signal received at the BS is a superposition of all users. Consequently, the preprocessing at the BS involves MMSE-based MUD and OFDM demapping, followed by converting the complex signal back to a real-valued tensor ${\mathbf{a}\in \mathbb{R}^{2Km}}$. Finally, $\mathbf{a}$ is reconstructed into the dimensions of ${N_c\times N_t\times 2KN_r}$ through an FC layer and a reshape operation.

Thereafter, in order to obtain the analog and digital beamforming matrices, the aforementioned output is fed into the DJSCC decoder, which comprises three parts. The input dimension of the first ${L_3}$ MAXIM blocks is ${N_c\times N_t\times 2KN_r}$, and the output is ${\mathbf{F}_{\mathrm{virt}}\in \mathbb{R}^{N_c\times N_t\times 2KN_s}}$, which serves as an intermediate virtual fully-digital beamforming representation. This tensor captures the global spatial-frequency features required for hybrid beamforming design. Taking ${\mathbf{F}_{\mathrm{virt}}}$ as a shared input, the decoder then branches into two parallel paths. The second ${L_4}$ MAXIM blocks generates an output of dimension ${N_c\times N_t\times N_{\mathrm{RF}}}$, which is then averaged along the dimension of ${N_c}$ to obtain the analog beamformer’s phase values ${\mathbf{\Theta }_{\mathrm{RF}}\in \mathbb{R}^{N_t\times N_{\mathrm{RF}}}}$. The third ${L_5}$ MAXIM blocks produces an output of dimension ${N_c\times N_t\times 2N_{\mathrm{RF}}KN_s}$, which is averaged along the dimension of ${N_t}$, and reshaped into the real-valued digital beamforming matrix ${\mathbf{F}_{\mathrm{BB}}\in \mathbb{R}^{N_c\times N_{\mathrm{RF}}\times 2KN_s}}$. Furthermore, $\mathbf{F}_{\mathrm{BB}}$ is converted into its complex-valued form $\mathbf{F}_{\mathrm{BB}} \in \mathbb{C}^{N_c \times N_{\mathrm{RF}} \times KN_s}$. Finally, the analog and digital beamforming matrices ${\mathbf{F}_{\mathrm{RF}}}$ and ${\mathbf{F}_{\mathrm{BB}}}$ are determined according to Eq. (\ref{Eq:max_sum_rate}), Eq. (\ref{Eq:RF_limit}) and Eq. (\ref{Eq:RFBB_limit}).

\begin{figure*}[t]
	\centering
	\includegraphics[width=0.85\textwidth]{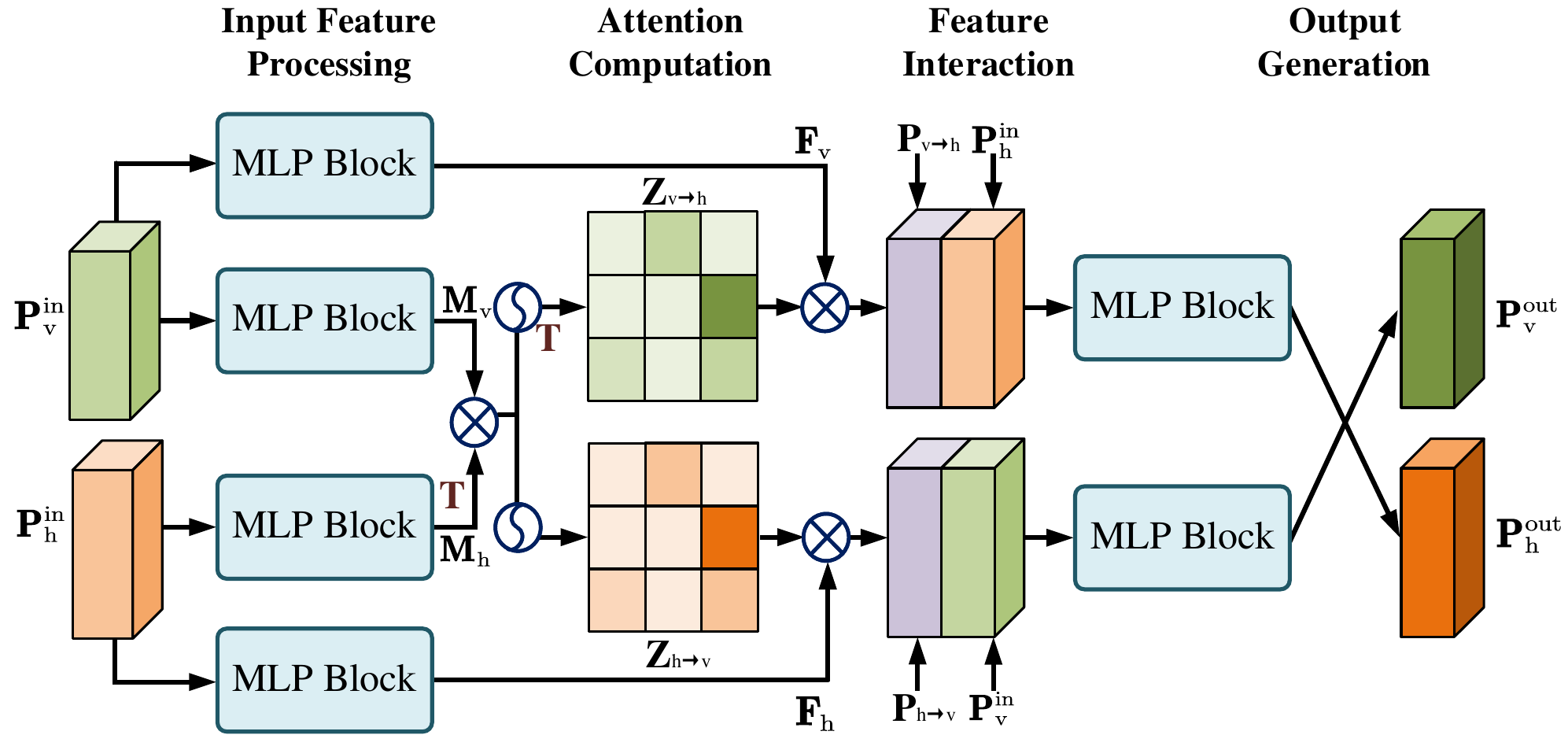}
	%\vspace{-15pt}
	\caption{Structure of the CPI module.}
	\vspace{-15pt}  % 减少图片下方的间距
	\label{fig:CPI module}
\end{figure*}

\subsection{Cross-Polarization Interaction Module for Joint Dual-Polarized CSI Compression}
Dual-polarized MIMO-OFDM systems typically utilize horizontal and vertical polarization modes to enhance spectral efficiency and channel capacity. The horizontal and vertical CSI components are characterized by significant structural correlation, as they share similar angles of arrival and multipath propagation delays due to the nature of dual-polarized propagation. While conventional CSI feedback schemes designed for single-polarized systems fail to exploit these dependencies, dual-polarized systems can leverage the cross-polarization correlations to facilitate joint compression and reconstruction. \cite{ying2020SAM} proposes a generic stereo attention module to calculate bidirectional stereo attention for stereo image super-resolution. Inspired by \cite{ying2020SAM}, we propose a CPI module for dual-polarized MIMO-OFDM systems that models the vertical and horizontal polarization CSI as two modalities and enables cross-polarization feature interaction via a co-attention mechanism. This approach enables the model to extract complementary information from both polarization components, thereby enhancing cross-polarization alignment. Consequently, the integrity of the channel representation is preserved and enriched during the compression phase, effectively mitigating distortion induced by compression.

As illustrated in Fig. \ref{fig:CPI module}, the input CSI features corresponding to horizontal and vertical polarizations, denoted as $\mathbf{P}_{\mathrm{h}}^{\mathrm{in}}, \mathbf{P}_{\mathrm{v}}^{\mathrm{in}} \in \mathbb{R}^{N_c \times N_t \times 2}$, are first reshaped into $N_c \times 2N_t$ by concatenating their real and imaginary components along the second dimension. This operation ensures the input format satisfies the requirements of the subsequent MLP layers for feature extraction. These inputs are initially processed by an MLP block to extract deep features, yielding the intermediate representations ${\mathbf{M}_{\mathrm{h}},\mathbf{M}_{\mathrm{v}}\in \mathbb{R}^{N_c\times 2N_t}}$. In the attention computation stage, to capture cross-polarization dependencies, we compute bidirectional attention maps. First, ${\mathbf{M}_{\mathrm{h}}}$ and ${\mathbf{M}_{\mathrm{v}}}$ are transposed to obtain ${\mathbf{M}_{\mathrm{h}}^{\mathrm{T}},\mathbf{M}_{\mathrm{v}}^{\mathrm{T}}\in \mathbb{R}^{2N_t\times N_c}}$. The similarity score maps are then calculated via matrix multiplication, followed by a column-wise softmax operation to produce the normalized attention maps ${\mathbf{Z}_{\mathrm{v}\rightarrow \mathrm{h}},\mathbf{Z}_{\mathrm{h}\rightarrow \mathrm{v}}\in \mathbb{R}^{N_c\times N_c}}$. The process of obtaining these bidirectional attention maps can be formulated as follows
\begin{equation}
	\label{Eq:Zv-h}
	\mathbf{Z}_{\mathrm{v}\rightarrow \mathrm{h}}=\mathrm{softmax}\left( \mathbf{M}_{\mathrm{h}}\mathbf{M}_{\mathrm{v}}^{\mathrm{T}} \right),
\end{equation}
\begin{equation}
	\label{Eq:Zh-v}
	\mathbf{Z}_{\mathrm{h}\rightarrow \mathrm{v}}=\mathrm{softmax}\left( \mathbf{M}_{\mathrm{v}}\mathbf{M}_{\mathrm{h}}^{\mathrm{T}} \right),
\end{equation}
where $\mathrm{softmax}(\cdot)$ denotes the column-wise softmax function. The matrix products ${\mathbf{M}_{\mathrm{h}}\mathbf{M}_{\mathrm{v}}^{\mathrm{T}}}$ and ${\mathbf{M}_{\mathrm{v}}\mathbf{M}_{\mathrm{h}}^{\mathrm{T}}}$ denote the cross-polarization similarity between the vertical and horizontal features. Accordingly, the normalized attention matrices $\mathbf{Z}_{\mathrm{v}\rightarrow \mathrm{h}}$ and $\mathbf{Z}_{\mathrm{h}\rightarrow \mathrm{v}}$ quantify the cross-polarization dependencies between the two polarization branches. Specifically, the element ${\left[ \mathbf{Z}_{\mathrm{v}\rightarrow \mathrm{h}}\right]_{i,j}}$ indicates the normalized attention weight corresponding to the similarity between the ${i}$-th row of vertical feature ${\mathbf{M}_\mathrm{h}}$ and the ${j}$-th column of transposed horizontal feature ${\mathbf{M}_\mathrm{v}^{\mathrm{T}}}$.

During the feature interaction stage, the attention maps facilitate the weighted aggregation of the intermediate features ${\mathbf{F}_{\mathrm{h}}, \mathbf{F}_{\mathrm{v}}\in\mathbb{R}^{N_c\times 2N_t}}$, thereby highlighting task-relevant information. These features are derived from the initial inputs  ${\mathbf{P}_{\mathrm{h}}^{\mathrm{in}}}$ and ${\mathbf{P}_{\mathrm{v}}^{\mathrm{in}}}$ via an MLP block. Formally, the cross-polarization feature interaction is expressed as
\begin{equation}
	\label{Eq:Pv-h}
\mathbf{P}_{\mathrm{v}\rightarrow \mathrm{h}}=\mathbf{Z}_{\mathrm{v}\rightarrow \mathrm{h}}\mathbf{F}_{\mathrm{v}},
\end{equation}
\begin{equation}
	\label{Eq:Ph-v}
\mathbf{P}_{\mathrm{h}\rightarrow \mathrm{v}}=\mathbf{Z}_{\mathrm{h}\rightarrow \mathrm{v}}\mathbf{F}_{\mathrm{h}},
\end{equation}
where ${\mathbf{P}_{\mathrm{v}\rightarrow \mathrm{h}}, \mathbf{P}_{\mathrm{h}\rightarrow \mathrm{v}}\in \mathbb{R}^{N_c\times 2N_t}}$ denote the cross-polarization features aligned from the vertical to the horizontal branch and from the horizontal to the vertical branch, respectively. To fuse these cross-polarization dependencies with the local characteristics, the interacted features are concatenated with their respective original inputs.
The final output features are then generated by an MLP block, which can be expressed as
\begin{equation}
	\label{Eq:Ph-out}
\mathbf{P}_{\mathrm{h}}^{\mathrm{out}}=\mathrm{MLP}\left( \mathrm{concat}\left( \mathbf{P}_{\mathrm{v}\rightarrow \mathrm{h}},\mathbf{P}_{\mathrm{h}}^{\mathrm{in}} \right) \right), 
\end{equation}
\begin{equation}
	\label{Eq:Pv-out}
\mathbf{P}_{\mathrm{v}}^{\mathrm{out}}=\mathrm{MLP}\left( \mathrm{concat}\left( \mathbf{P}_{\mathrm{h}\rightarrow \mathrm{v}},\mathbf{P}_{\mathrm{v}}^{\mathrm{in}} \right) \right), 
\end{equation}
where the output features ${\mathbf{P}_{\mathrm{h}}^{\mathrm{out}}, \mathbf{P}_{\mathrm{v}}^{\mathrm{out}}\in \mathbb{R}^{N_c\times N_t\times 2}}$ are reshaped back to their original dimensions to maintain structural consistency with the original input CSI.

\section{Simulation Results}
In this section, we compare the proposed scheme with existing DL-based and conventional methods considering different numbers of users and various uplink and downlink signal-to-noise ratio (SNR) conditions, demonstrating the effectiveness of the proposed scheme. Furthermore, we verify the superiority of the DJSCC-based feedback strategy adopted in our scheme over the conventional SSCC-based approach. Finally, we validate the effectiveness of the proposed CPI module through ablation studies.

\begin{figure*}[t]
	\vspace{-9mm}
	\captionsetup{font={footnotesize, color = {black}}, name = {Fig.}, labelsep = period} % singlelinecheck = off, justification = raggedright,
	\captionsetup[subfigure]{singlelinecheck = on, justification = raggedright, font={footnotesize}}
	\centering
	\hspace{-3.0mm}
	\subfloat[]{%[Training overhead $T_{\rm{CE}} = 88$]
		\label{fig:4a}
		\begin{minipage}[t]{0.48\linewidth}
			\centering
			\includegraphics[scale=0.5]{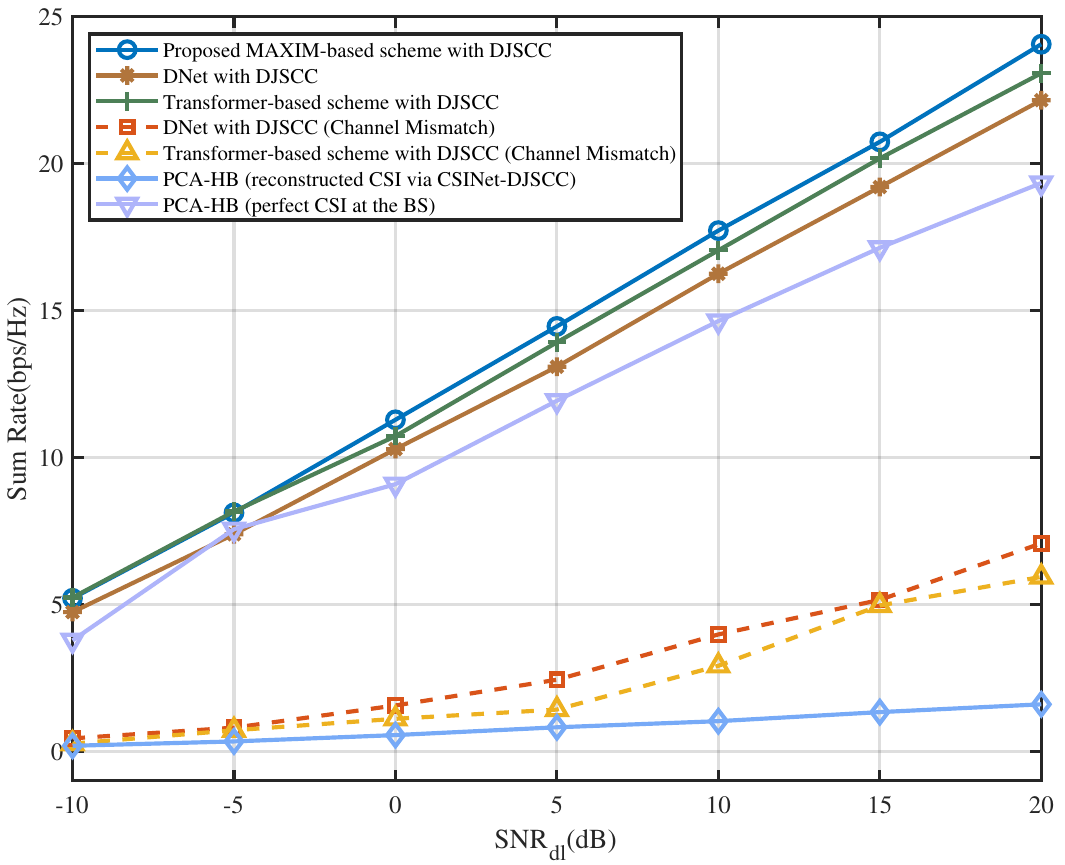}
			%			\vspace{-10.0mm}
		\end{minipage}
	}
	\hfill
	\subfloat[]{%[Training overhead $T_{\rm{CE}} = 176$]
		\label{fig:4b}
		\begin{minipage}[t]{0.48\linewidth}
			\centering
			\includegraphics[scale=0.5]{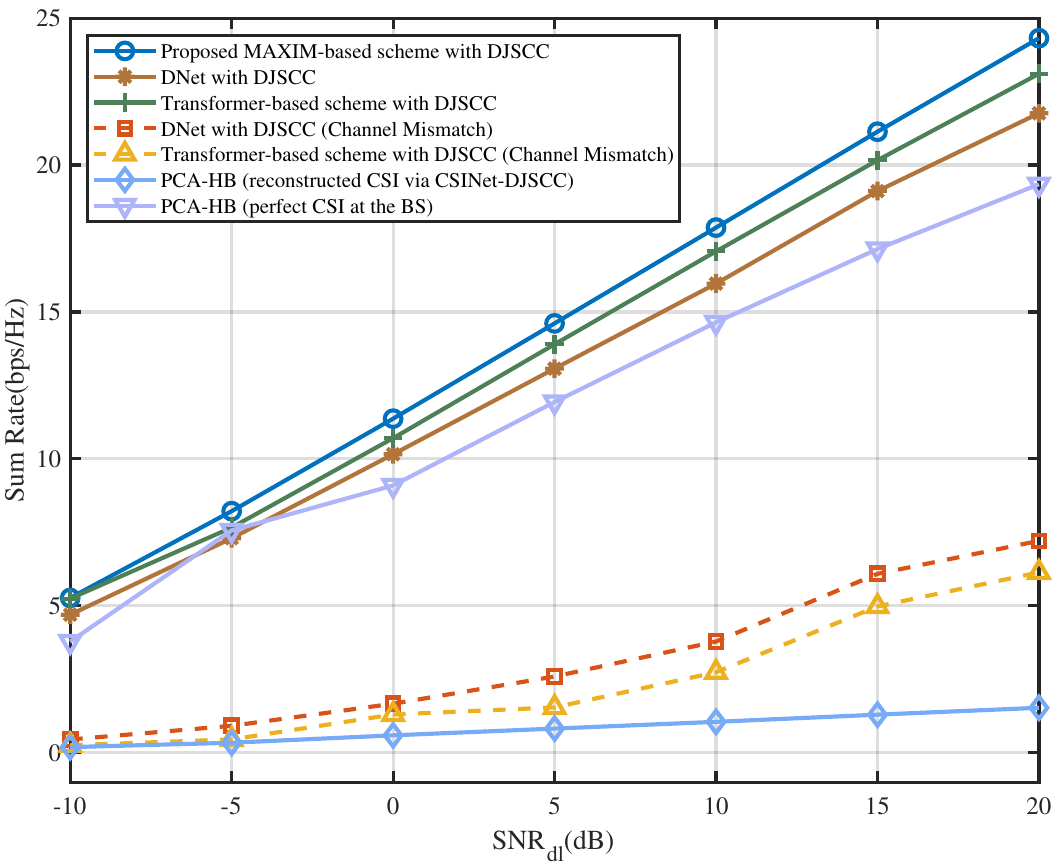}
			%			\vspace{-10.0mm}
		\end{minipage}
	}
	\vspace{-1.0mm}
	%	\caption{NMSE performance comparison of different CE schemes versus SNRs with the same number of
		%		MPCs $L=4$: (a)~training overhead $T_{\rm CE}=88$; and (b) $T_{\rm CE}=176$.}
	\caption{Sum rate achieved by different schemes versus downlink SNR, given $K=2$ and $m=10$ : (a) $\mathrm{SNR}_\mathrm{ul}=0$ dB; (b) $\mathrm{SNR}_\mathrm{ul}=20$ dB.}
	\label{fig:vsbaseline}
	\vspace{-5.0mm}
\end{figure*}

We consider a typical 6G cmWave urban macrocell (UMa) scenario where the downlink and uplink center frequencies are set to 6.7 GHz and 6 GHz, respectively. The uplink and downlink channel datasets are generated using QuaDRiGa \cite{jaeckel2014quadriga} based on the 3GPP TR 38.901 specifications \cite{3gpp2017study}. The OFDM system employs ${N_c=32}$ subcarriers. The BS is equipped with ${N_t=64}$ antenna ports, and each UE is equipped with ${N_r=2}$ antenna ports, corresponding to 32 dual-polarized antennas at the BS and one dual-polarized antenna at each UE. We assume that the number of users ${K}$ equals the number of RF chains ${N_{\mathrm{RF}}}$. In the subsequent simulations, we consider scenarios with ${K=2}$ and ${K=4}$, respectively. Additionally, the number of data streams per user is set to ${N_s=1}$. The UEs are randomly distributed within a ${20m\times20m}$ square region. The embedding dimension ${d_{model}}$ in MAXIM is set to 256. For the MAXIM block configuration, we set the number of layers as $L_1=L_3=1$ and $L_2=L_4=L_5=2$.

Additionally, we use the open-source DL library PyTorch to train and validate the proposed scheme on a computer with dual Nvidia GeForce GTX 4090D GPUs. The training, validation and testing datasets contain 90,000, 10,000, and 10,000 samples, respectively. Each uplink and downlink CSI sample is power-normalized. The batch size is 256 and the number of training epochs is 1,000. We employ the Adam optimizer to train the entire neural network in an end-to-end manner, with the learning rate scheduled by the NoamOpt scheduler \cite{vaswani2017attention}. Furthermore, the average user sum rate defined in Eq. (\ref{Eq:sum_rate}) serves as the loss function.

\subsection{Performance Comparison with Baseline Methods}
In our experiments, the subcarrier overhead is fixed at $m=10$ for all considered schemes. We evaluate the achievable downlink sum rate versus downlink SNR over the range of [-10,20] dB, respectively given uplink SNRs of 0 dB and 20 dB. To ensure a fair comparison, all DJSCC-based baselines (including DNet, transformer, and PCA-HB with reconstructed CSI) share a common uplink transmission framework. Specifically, they adopt a TDMA protocol where users feed back their CSI sequentially, and the BS employs maximal ratio combining (MRC) to recover individual signals, resulting in a total overhead of $K$ time slots. To ensure a fair comparison, the proposed scheme and baselines are compared under the same total feedback payload of $Km$ symbols transmitted to the BS. Accordingly, the TDMA schemes occupy $K$ time slots with $m$ subcarriers each, resulting in a total of $Km$ uplink time-frequency resource units. In contrast, the proposed scheme uses only $m$ uplink resource units, which are simultaneously shared by all $K$ users. The specific schemes are summarized as follows:

\begin{itemize}
	\item \textbf{Proposed MAXIM-based scheme with DJSCC}: Unlike the TDMA baselines, this scheme enables simultaneous CSI feedback for all $K$ users within only one time slot. Specifically, all users simultaneously employ DJSCC for CSI feedback over the realistic uplink MIMO-OFDM channel, and the BS utilizes MMSE-based MUD to decouple the superimposed signals. The entire framework is trained end-to-end to maximize the achievable downlink sum rate.
\end{itemize}

\begin{itemize}
	\item \textbf{DNet with DJSCC}: DNet is a CNN-based approach originally proposed in \cite{wei2022distributed}, where feedback errors are modeled by directly adding noise to the input of the decoder. For this baseline, we adapt the DNet architecture to facilitate CSI feedback via DJSCC over a realistic uplink MIMO-OFDM channel, incorporating the TDMA-MRC protocol.	
\end{itemize}

\begin{itemize}
	\item \textbf{Transformer-based scheme with DJSCC}: This baseline is based on the transformer-based hybrid beamforming scheme proposed in \cite{wang2022transformer}. While the original work considers only quantization errors without modeling channel impairments, we adapt its architecture for DJSCC-based CSI feedback over a realistic uplink MIMO-OFDM channel under the TDMA-MRC framework.
\end{itemize}

\begin{itemize}
	\item \textbf{DNet with DJSCC (Channel Mismatch)}: Unlike DNet with DJSCC, this variant is trained under an AWGN channel environment but evaluated within the realistic MIMO-OFDM environment. This channel mismatch configuration investigates the performance degradation caused by the discrepancy between the channel statistics assumed during training and those actually encountered during deployment.
\end{itemize}

\begin{itemize}
	\item \textbf{Transformer-based scheme with DJSCC (Channel Mismatch)}:	Adopting the same configuration, this variant assesses the impact of channel mismatch specifically on the transformer-based architecture.
\end{itemize}

\begin{itemize}
	\item \textbf{PCA-HB (reconstructed CSI via CSINet-DJSCC)}: This scheme first reconstructs multi-user uplink CSI using a DJSCC-based CSINet under the TDMA-MRC protocol, followed by the PCA-based hybrid beamforming algorithm from \cite{sun2020principal}. The normalized mean square error is adopted as the loss function to train the DJSCC-based CSINet.
\end{itemize}
		
\begin{itemize}
	\item \textbf{PCA-HB (perfect CSI at the BS)}: This scheme assumes that perfect downlink CSI is available at the BS, thereby eliminating the need for CSI feedback. Based on this perfect CSI, the BS performs  hybrid beamforming using the PCA-based algorithm proposed in \cite{sun2020principal}.
\end{itemize}

\begin{figure*}[t]
	\vspace{-9mm}
	\captionsetup{font={footnotesize, color = {black}}, name = {Fig.}, labelsep = period} % singlelinecheck = off, justification = raggedright,
	\captionsetup[subfigure]{singlelinecheck = on, justification = raggedright, font={footnotesize}}
	\centering
	\hspace{-3.0mm}
	\subfloat[]{%[Training overhead $T_{\rm{CE}} = 88$]
		\label{fig:5a}
		\begin{minipage}[t]{0.48\linewidth}
			\centering
			\includegraphics[scale=0.49]{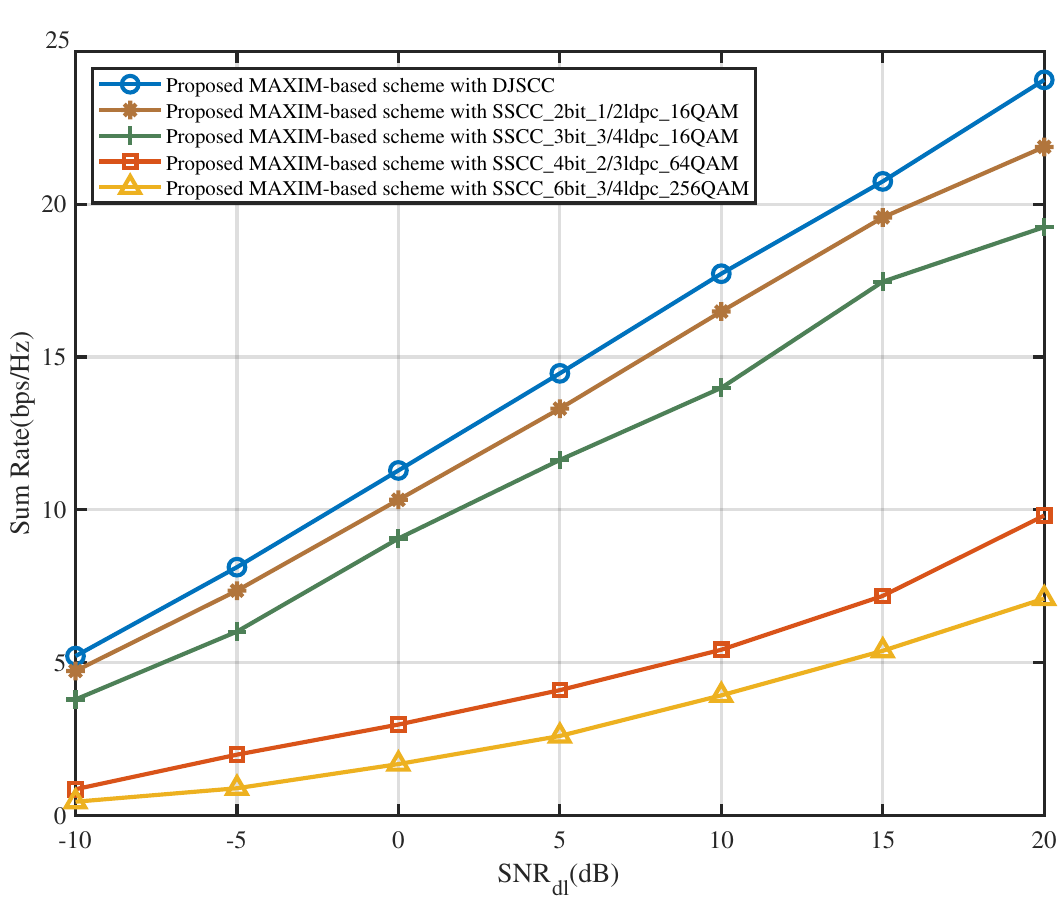}
			%			\vspace{-10.0mm}
		\end{minipage}
	}
	\hfill
	\subfloat[]{%[Training overhead $T_{\rm{CE}} = 176$]
		\label{fig:5b}
		\begin{minipage}[t]{0.48\linewidth}
			\centering
			\includegraphics[scale=0.49]{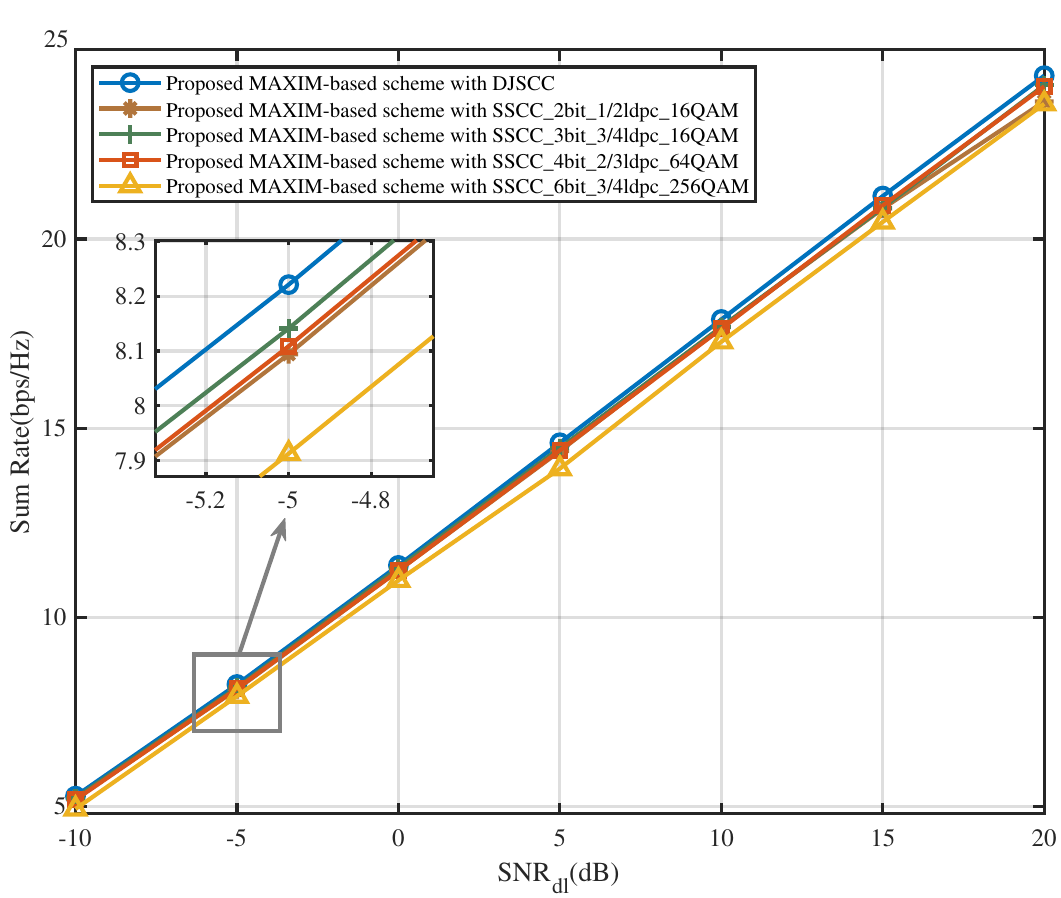}
			%			\vspace{-10.0mm}
		\end{minipage}
	}
	\vspace{-1.0mm}
	\caption{Sum-rate performance comparison of the proposed DJSCC-based multi-user network and conventional SSCC-based schemes, given $K=2$ and $m=10$ : (a) $\mathrm{SNR}_\mathrm{ul}=0$ dB; (b) $\mathrm{SNR}_\mathrm{ul}=20$ dB.}
	\label{fig:vsSSCC}
	\vspace{-5.0mm}
\end{figure*}

Fig. \ref{fig:vsbaseline} illustrate the downlink achievable sum rate of the proposed scheme versus the baseline schemes across varying SNR conditions. It can be observed that the proposed scheme consistently outperforms the baseline schemes. Notably, the PCA-HB scheme utilizing DJSCC-based CSINet for CSI reconstruction suffers from significant performance degradation due to the low dimension of the DJSCC encoder output, which causes CSI reconstruction failure and subsequent hybrid beamforming misalignment. These results suggest that, under limited subcarrier overhead, the proposed end-to-end direct mapping strategy is more effective than the conventional two-stage reconstruction-and-design approach for hybrid beamforming. The performance gain arises from adopting a task-oriented design rather than a data-oriented one. Specifically, by avoiding explicit CSI reconstruction, the proposed method reduces the mismatch between the intermediate reconstruction objective and the final beamforming task, enabling the network to focus on channel features that are most relevant to hybrid beamforming design. Compared to DNet with DJSCC and transformer-based scheme with DJSCC, their corresponding “Channel Mismatch” variants exhibit significant performance degradation. Specifically, at $\mathrm{SNR}_\mathrm{ul} = 0$ dB, the sum rate decreases by 4.29–15.08 bps/Hz for the DNet variant and by 4.97–17.14 bps/Hz for the transformer-based variant. These results demonstrate that models trained under ideal channel conditions are not suitable for realistic communication scenarios.

Moreover, it is observed that the proposed MAXIM-based scheme with DJSCC consistently outperforms the perfect-CSI PCA-HB baseline, with the performance advantage becoming more pronounced as the downlink SNR increases. This indicates that the PCA-HB baseline, even with perfect CSI, does not represent the performance upper bound under the same hybrid architecture constraints. This performance gap stems from the fundamental difference in optimization paradigms. Conventional heuristic algorithms like PCA-HB inherently rely on a decoupled optimization approach, where the analog beamformer is designed via singular value decomposition (SVD) to maximize directional gain and the digital precoder is subsequently designed via zero-forcing (ZF) for interference nulling. Such decoupled designs often fail to achieve the global sum-rate maximum due to the lack of joint optimization between the analog and digital stages. In contrast, our proposed MAXIM-based DJSCC framework treats the entire transmission chain as a unified system, directly optimizing the hybrid beamforming matrices with the multi-user sum rate as the objective in an end-to-end manner. By learning task-oriented channel representations, this joint optimization is capable of exploring more effective precoding strategies that overcome the limitations of conventional decoupled heuristic rules. This advantage remains consistent even when the PCA-HB baseline is provided with perfect CSI, whereas our scheme is constrained by limited feedback and the resulting distortion. Although the TDMA-MRC protocol employed in the DNet and transformer-based schemes circumvents multi-user interference through temporal orthogonality, it relies on a fundamentally noise-limited technique that lacks the capability of suppressing inter-user interference. This inherent limitation forces such frameworks to adopt sequential feedback, requiring $K$ time slots and resulting in high latency. As the number of users increases, the resulting feedback overhead further exacerbates channel aging, which reduces CSI reliability and limits beamforming performance in time-varying environments. In contrast, the proposed MAXIM-based DJSCC scheme leverages simultaneous multi-user feedback and MMSE-based MUD. This approach effectively exploits the spatial structure of the uplink MIMO channel for more precise CSI recovery while preserving multi-user joint characteristics. Additionally, the proposed scheme enables simultaneous feedback within a single time slot, which is crucial for meeting the ultra-low latency requirements of mission-critical IIoT applications, such as real-time industrial automation and robotic control. By reducing the transmission latency from $K$ time slots to 1, this mechanism inherently mitigates channel aging, ensuring that the semantic features received at the BS remain representative of the current channel state. Driven by end-to-end sum-rate optimization, the proposed MAXIM-based DJSCC scheme establishes a superior synergistic design among feedback, recovery, and beamforming, thus outperforming the schemes based on the TDMA-MRC protocol. Specifically, at uplink and downlink SNRs of 20 dB, the proposed scheme achieves sum-rate gains of 2.56 bps/Hz and 1.21 bps/Hz over the DNet with DJSCC and the transformer-based scheme with DJSCC, respectively. Moreover, unlike conventional reconstruction-based approaches that are sensitive to feedback latency, our task-oriented design focuses on extracting spatial correlations essential for beamforming, which exhibit stronger resilience against feedback delays. Specifically, while instantaneous channel phases fluctuate rapidly, the underlying spatial correlations, such as the angular power spectrum, characterize the slowly-evolving structural properties of the environment. By prioritizing these stable semantic features, the framework maintains the essential directional information required for beamforming, thereby mitigating the performance degradation caused by channel aging.

\subsection{Performance Comparison with SSCC Feedback Schemes}

Fig. \ref{fig:vsSSCC} compares the sum rate of the proposed DJSCC-based multi-user joint CSI feedback and hybrid beamforming network with conventional SSCC schemes under the case of 2 UEs. In the DJSCC scheme, the real-valued encoder outputs are converted into complex-valued symbols for transmission over the uplink MIMO channel. At the BS, these signals are converted back into real-valued representations before being fed into the decoder. In contrast, the SSCC schemes quantize the real-valued encoder outputs uniformly to generate bit streams. Following the methodology in \cite{zhang2025djscc}, we assume perfect bit-level feedback during the training phase. In the testing phase, practical transmission over the uplink MIMO channel is implemented using low-density parity-check (LDPC) coding and quadrature amplitude modulation (QAM), followed by uniform de-quantization at the BS.

To ensure a fair comparison, the number of feedback symbols is kept identical for both DJSCC and SSCC schemes. For the DJSCC scheme, a feedback overhead of $d_{k} = 10$ symbols is allocated to each user. Consequently, the total symbol overhead for a two-user simultaneous uplink scenario is 20. For the SSCC schemes, we denote the configuration as $\mathrm{SSCC}\_q\mathrm{bit\_}r\mathrm{ldpc\_}a\mathrm{QAM}$, where each element of the encoder output is quantized to $q$ bits, followed by LDPC coding with rate $r$ and QAM modulation with $a$ constellation points. Consequently, the number of uplink feedback symbols per user in the SSCC scheme is determined by $d_{k} = \frac{b \times q}{r \times \log_2 a}$, where $b$ represents the dimension of the encoder output.

In Fig. \ref{fig:5a} and Fig. \ref{fig:5b}, the performance of the proposed DJSCC scheme consistently outperforms the SSCC schemes across all downlink SNRs, demonstrating the superiority of our DJSCC scheme. As illustrated in Fig. \ref{fig:5a}, the DJSCC scheme exhibits a more significant performance gain compared to the SSCC schemes under unfavorable channel conditions, achieving a sum rate gain of 4.76–16.96 bps/Hz over the $\mathrm{SSCC}\_6\mathrm{bit\_}3/4\mathrm{ldpc\_}256\mathrm{QAM}$ scheme. As depicted in Fig. \ref{fig:5b}, the performance gap between the SSCC and DJSCC schemes narrows under favorable channel conditions. Nevertheless, the DJSCC scheme consistently outperforms the SSCC schemes, achieving a gain of 0.31–0.75 bps/Hz compared to the $\mathrm{SSCC}\_6\mathrm{bit\_}3/4\mathrm{ldpc\_}256\mathrm{QAM}$ scheme. Furthermore, Fig. \ref{fig:5a} reveals that the 1/2-rate LDPC SSCC scheme outperforms SSCC schemes of other rate configurations under unfavorable channel conditions. This is primarily because a lower code rate enhances transmission reliability against severe channel impairments, which in turn facilitates more accurate CSI feedback and improves hybrid beamforming performance of the overall system.
\begin{figure}[t]
	\centering
	\includegraphics[width=0.49\textwidth]{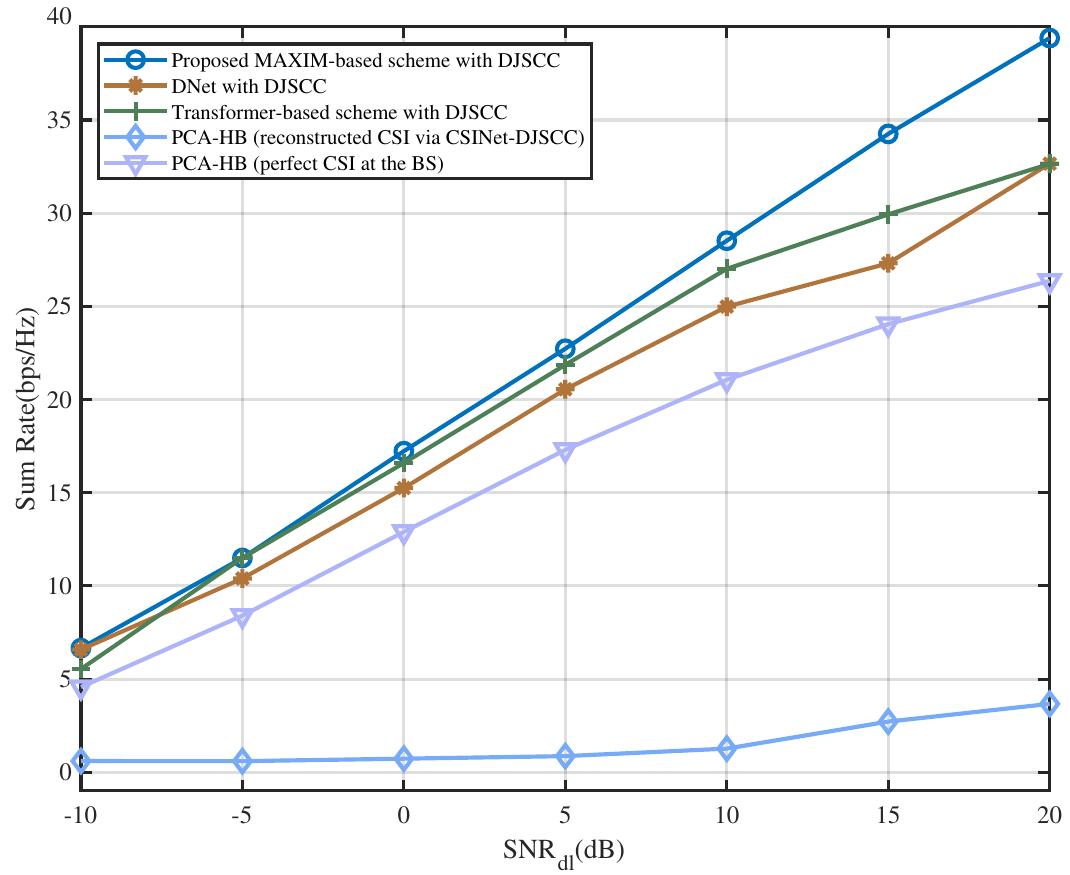}
	%\vspace{-15pt}
	\caption{Sum rate achieved by different schemes 
		versus downlink SNR at $\mathrm{SNR}_\mathrm{ul}=0$ dB, given $K=4$ and $m=20$.}
	\vspace{-15pt}  % 减少图片下方的间距
	\label{fig:vsdjscc_4user}
\end{figure}

\renewcommand{\arraystretch}{1.2}

\begin{table*}[htbp]
	\centering
	\captionsetup{labelfont={color=red}}
	\caption{{Sum-rate performance comparison of the proposed scheme with and without the CPI module versus downlink SNR (for ${K=2}$ and $\mathrm{SNR}_\mathrm{ul}=0\text{dB}$)}}
	\begin{tabular}{c|c|c|c|c|c|c|c}  % 所有列设置为居中对齐
		\hline
		$\text{SNR}_{\mathrm{dl}}$ & -10 dB & -5 dB & 0 dB & 5 dB & 10 dB & 15 dB & 20 dB  \\ \hline
		\textbf{Proposed w/o CPI} & 5.214±0.025 & 8.122±0.030 & 11.270±0.037 & 14.466±0.047 & 17.582±0.062 & 20.020±0.071 & 23.636±0.096 \\ 
		\textbf{Proposed w/ CPI} & 5.228±0.025 & 8.140±0.022 & 11.288±0.031 & 14.482±0.032 & 17.726±0.047 & 20.746±0.061 & 24.068±0.081\\ \hline
		
	\end{tabular}
	\label{tb:CPI_Ablation}
\end{table*}

\begin{comment}
\begin{table*}[!t]
	\centering
	\captionsetup{labelfont={color=red}}
	{
		\caption{{Sum-rate Performance Comparison of the Proposed Scheme with and without the CPI Module versus Downlink SNR (for ${K=2}$ and $\mathrm{SNR}_\mathrm{ul}=0\text{dB}$)}}
		\setlength{\tabcolsep}{9pt} % 默认列间距设置为12pt
		\begin{tabular}{c|c|c|c|c|c|c|c}  % 所有列设置为居中对齐
			\hline
			$\text{SNR}_{\mathrm{dl}}$ & -10 dB & -5 dB & 0 dB & 5 dB & 10 dB & 15 dB & 20 dB  \\ \hline
			\textbf{Proposed w/o CPI} & 5.213951 & 8.123524 & 11.270971 & 14.465367 & 17.582334 & 20.019758 & 23.635221 \\ 
			\textbf{Proposed w/ CPI} & 5.228202 & 8.140897 & 11.288336 & 14.483163 & 17.725811 & 20.745985 & 24.067848\\ \hline
			
		\end{tabular}
		\vspace{-10pt}  % 减少图片下方的间距
		\label{tb:CPI_Ablation}
	}
\end{table*}
\end{comment}

\subsection{Multi-User Performance Comparison of the Proposed and Baseline Methods}
To evaluate the scalability of the proposed framework, Fig. \ref{fig:vsdjscc_4user} compares the sum-rate performance of various schemes in a multi-user scenario with $K=4$ users. In these simulations, the subcarrier overhead is fixed at $m=20$ for all schemes. It can be observed that the proposed MAXIM-based scheme with DJSCC consistently outperforms all other baselines across the entire downlink SNR range. Notably, the performance gap between the proposed scheme and other DL-based schemes expands as the downlink SNR increases. At $\mathrm{SNR}_\mathrm{dl} = 20$ dB, the proposed scheme achieves a sum rate of approximately 40 bps/Hz, outperforming the transformer-based and DNet-based schemes by 6.75 bps/Hz, demonstrating its superior capability in capturing and exploiting the underlying channel features for efficient hybrid beamforming under limited feedback overhead. Despite the reliance on perfect CSI in the conventional PCA-HB scheme, the proposed framework achieves a superior sum-rate performance. This suggests that the end-to-end learning paradigm of the DJSCC framework is more effective at maximizing the sum rate than the conventional hybrid beamforming modular designs, particularly when the available feedback resources are extremely limited. The observation that the conventional reconstruction-based scheme performs poorly in harsh scenarios indicates that the accuracy of reconstructed CSI cannot be guaranteed when feedback is severely degraded. Such results validate the necessity of our task-oriented DJSCC design, which prioritizes the end-to-end maximization of system performance over the intermediate accuracy of CSI reconstruction.

\subsection{Performance Comparison of the Proposed Scheme with and without the CPI Module}

To validate the effectiveness of the proposed CPI module, an ablation study is conducted as summarized in Table \ref{tb:CPI_Ablation}. The table compares the sum-rate performance of the proposed scheme with and without the CPI module across varying downlink SNR levels, given $K=2$ and $\mathrm{SNR}_\mathrm{ul}=0$dB. To improve the statistical reliability of the comparison, each result is averaged over $N=10$ independent runs, and the corresponding standard deviation is also reported. It can be observed that the scheme incorporating the CPI module consistently outperforms the baseline scheme without the CPI module over the entire SNR range. Although the absolute gain is relatively small in the low-SNR region, the improvement remains consistently positive across repeated runs, and becomes more evident in the medium-to-high SNR region. Specifically, at $\mathrm{SNR}_\mathrm{dl}=15$dB, the proposed scheme achieves a sum rate of $20.75$ bps/Hz compared with 20.02 bps/Hz without the CPI module. The additional variance analysis confirms that this improvement is stable rather than caused by random training fluctuations. This performance gain confirms that the CPI module plays a critical role in capturing cross-polarization features and mitigating compression-induced distortion, thereby effectively boosting the achievable sum rate.

\renewcommand{\arraystretch}{1.2}
\begin{table*}[htbp]
	\centering
	\captionsetup{justification=centering}
	\caption{{Sum-rate performance comparison for different subcarriers and downlink SNRs (for $K=2$ and $\mathrm{SNR_{ul}}=0\text{dB}$)}}
	\setlength{\tabcolsep}{9pt} % 默认列间距设置为12pt
	\begin{tabular}{c|c|c|c|c|c|c|c}  % 所有列设置为居中对齐
		\hline
		$\mathrm{SNR_{dl}}$ & -10 dB & -5 dB & 0 dB & 5 dB & 10 dB & 15 dB & 20 dB \\ \hline
		
		\textbf{16 subcarriers} & 5.179 & 8.076 & 11.187 & 14.350 & 17.509 & 20.781 & 23.553 \\ 
		
		\textbf{32 subcarriers} & 5.214 & 8.124 & 11.288 & 14.465 & 17.726 & 20.746 & 24.068 \\ 
		
		\textbf{64 subcarriers} & 5.251 & 8.141 & 11.306 & 14.474 & 17.725 & 20.962 & 23.784 \\ \hline
		
	\end{tabular}
	% \vspace{-18pt}  % 减少图片下方的间距
	\label{tb:different_subcarriers}
\end{table*}

\renewcommand{\arraystretch}{1.2}
\begin{table*}[htbp]
	\centering
	\captionsetup{justification=centering}
	\caption{{Sum-rate performance comparison at different downlink center frequencies and downlink SNRs (for $K=2$ and $\mathrm{SNR_{ul}}=0\text{dB}$)}}
	\setlength{\tabcolsep}{9pt} % 默认列间距设置为12pt
	\begin{tabular}{c|c|c|c|c|c|c|c}  % 所有列设置为居中对齐
		\hline
		$\mathrm{SNR_{dl}}$ & -10 dB & -5 dB & 0 dB & 5 dB & 10 dB & 15 dB & 20 dB \\ \hline
		
		\textbf{3GHz} & 5.232 & 8.139 & 11.276 & 14.454 & 17.704 & 20.327 & 23.408 \\ 
		
		\textbf{6.7GHz} & 5.214 & 8.124 & 11.288 & 14.465 & 17.726 & 20.746 & 24.068 \\ 
		
		\textbf{15GHz} & 5.240 & 8.121 & 11.282 & 14.418 & 17.535 & 20.619 & 24.208 \\ \hline
		
	\end{tabular}
	\vspace{-18pt}  % 减少图片下方的间距
	\label{tb:different_Downlink_Center_Frequencies}
\end{table*}

\subsection{Robustness and Adaptability Analysis}
To evaluate the robustness of the proposed MAXIM-based architecture for diverse IoT bandwidths, we conduct an experiment with different numbers of subcarriers $N_c$. As the number of subcarriers increases, the channel exhibits higher frequency selectivity. Table \ref{tb:different_subcarriers} compares the sum-rate performance for $N_c \in \{16, 32, 64\}$. The results indicate that the proposed framework remains effective across different frequency selectivity levels, with a maximum performance variation of less than $2.2\%$ at $\mathrm{SNR}_\mathrm{dl} = 20\mathrm{dB}$, thereby validating the robustness of the proposed design.
	
As illustrated in Table \ref{tb:different_Downlink_Center_Frequencies}, the proposed MAXIM-based scheme maintains a remarkably stable sum-rate performance across the carrier frequencies from 3 GHz to 15 GHz. Across the evaluated range, the sum-rate fluctuations remain within a narrow margin of less than $3.5\%$. Therefore, we can observe the robustness of the proposed scheme to different downlink center frequencies.

\begin{figure}[t]
	\centering
	\includegraphics[width=0.48\textwidth]{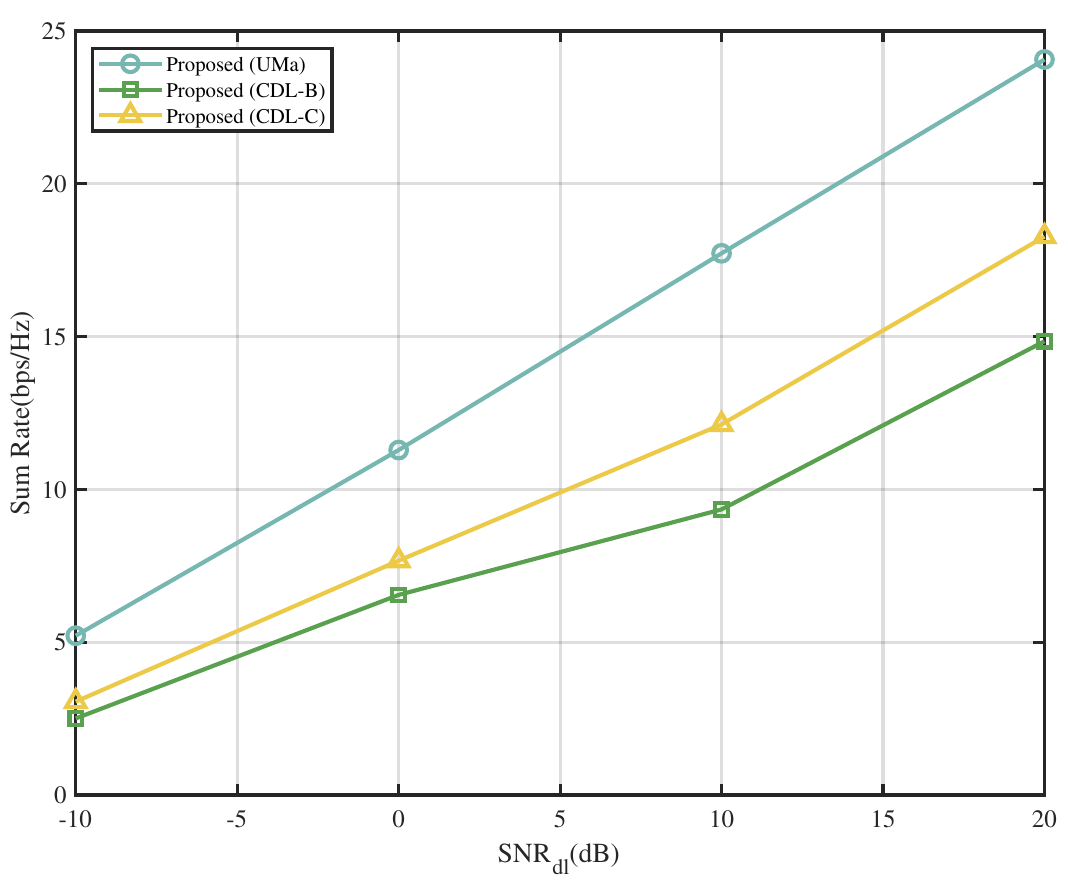}
	%\vspace{-15pt}
	\caption{Sum rate achieved by the  proposed MAXIM-based scheme across different channel environments at $\mathrm{SNR}_\mathrm{ul}=0\text{dB}$, given $K=2$ and $m=10$.}
	\vspace{-15pt}  % 减少图片下方的间距
	\label{fig:maxim_djscc_channel_environment}
\end{figure}

To investigate the robustness of the proposed architecture across different channel environments, we further evaluate the MAXIM-DJSCC framework under three representative 3GPP channel environments: UMa, clustered delay line (CDL)-B, and CDL-C. The results in Fig. \ref{fig:maxim_djscc_channel_environment} show that the framework maintains stable performance across these channel environments. Although CDL-B and CDL-C involve more challenging NLOS propagation characteristics, the framework remains effective, demonstrating its robustness and generalization capability under diverse scattering environments.

\begin{figure}[t]
	\centering
	\includegraphics[width=0.48\textwidth]{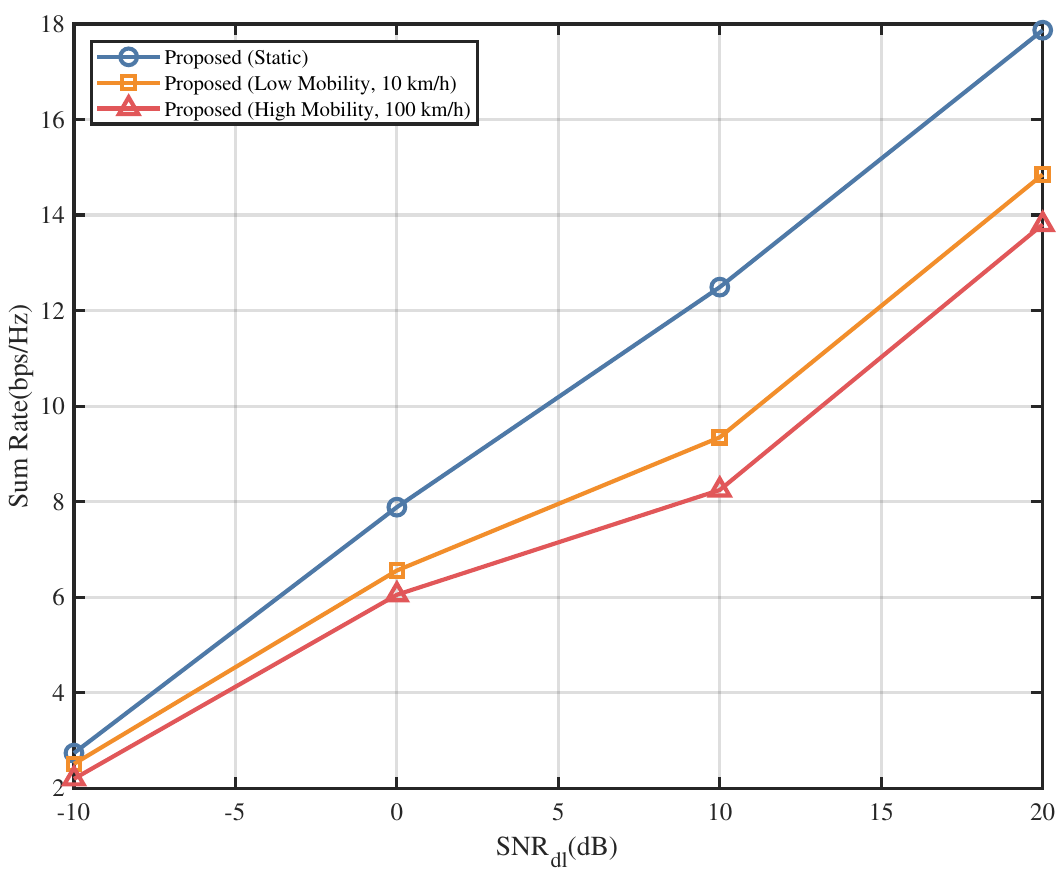}
	%\vspace{-15pt}
	\caption{Sum rate achieved by the proposed MAXIM-based scheme under the CDL-C channel environment across different mobility levels at $\mathrm{SNR}_\mathrm{ul}=0$ dB, given $K=2$ and $m=10$.}
	% \vspace{-15pt}  % 减少图片下方的间距
	\label{fig:maxim_djscc_mobility}
\end{figure}

Fig. \ref{fig:maxim_djscc_mobility} evaluates the sum-rate performance in the CDL-C channel environment under different UE velocities. As the UEs' velocity increases, the sum rate gradually decreases due to the more severe Doppler effect and the resulting channel aging. Nevertheless, the proposed scheme maintains a robust sum rate of $13.80\text{bps/Hz}$ at $100\text{ km/h}$ with $\mathrm{SNR}_\mathrm{dl}=20$dB, demonstrating the robustness of the MAXIM backbone against mobility-induced performance degradation even in the absence of a line-of-sight (LOS) component.

\begin{figure}[t]
	\centering
	\includegraphics[width=0.48\textwidth]{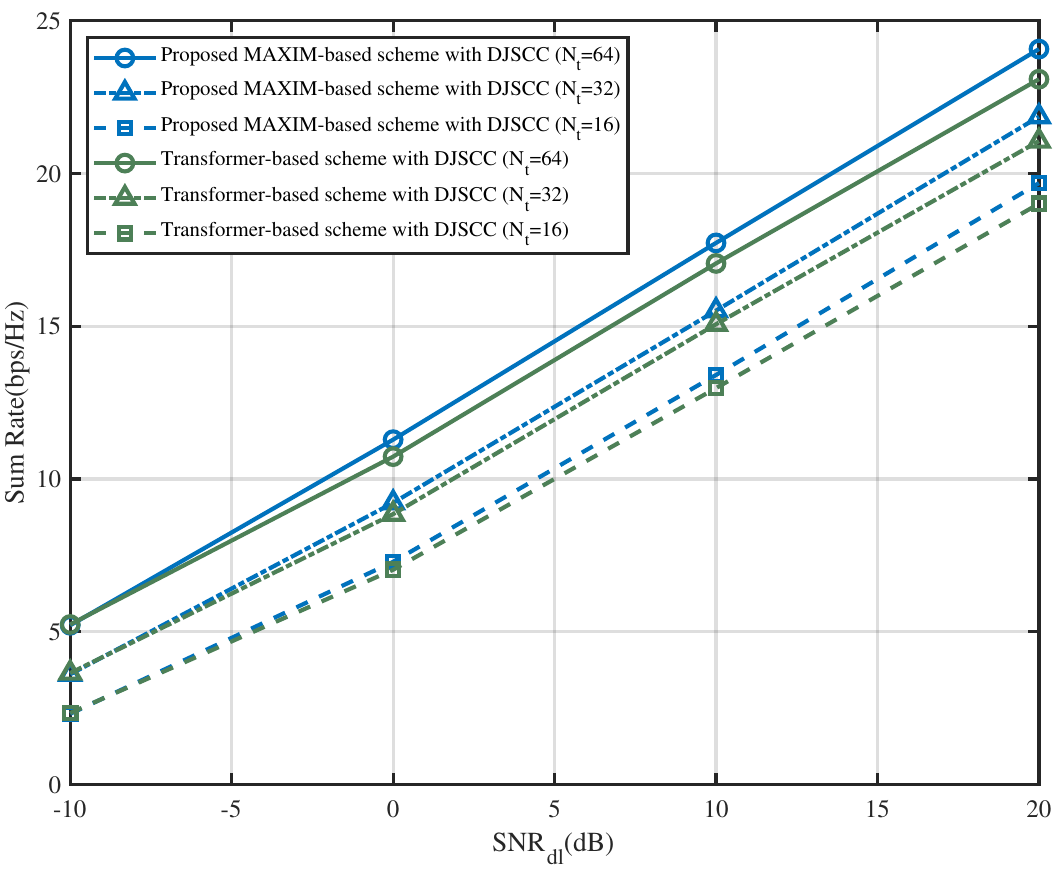}
	%\vspace{-15pt}
	\caption{Sum-rate performance versus downlink SNR for different antenna configurations ($N_t \in \{16, 32, 64\}$) at $\mathrm{SNR}_\mathrm{ul}=0$ dB, given $K=2$ and $m=10$.}
	\vspace{-15pt}  % 减少图片下方的间距
	\label{fig:maxim_transformer_djscc_tx}
\end{figure}

Fig. \ref{fig:maxim_transformer_djscc_tx} shows the performance comparison between the proposed MAXIM-based scheme and the transformer-based baseline under $N_t\in\{16, 32, 64\}$. As illustrated in Fig. \ref{fig:maxim_transformer_djscc_tx}, while both schemes benefit from the increased spatial degrees of freedom, the performance gain of proposed MAXIM-based scheme becomes more pronounced as the number of antennas increases. This demonstrates that the gated multi-axis interaction in MAXIM enables more robust and scalable feature extraction for high-dimensional CSI compared to self-attention-based backbones.

\subsection{Computational Complexity Analysis}
Table \ref{tab:complexity} compares the model parameters, floating-point operations (FLOPs), and inference latency of different DJSCC schemes and the conventional PCA-HB scheme, where the latter assumes perfect CSI. The proposed MAXIM model contains only 10.54 M parameters, which is the smallest among all compared DL-based schemes. This suggests that the proposed model is more parameter-efficient, achieving competitive performance with lower storage overhead. Although the FLOPs are relatively higher in the current configuration, the MAB within the proposed MAXIM-based schemes utilizes an axis-wise decomposition of interactions with a complexity of $\mathcal{O}(N_c N_t (N_c + N_t))$, thereby avoiding the quadratic complexity $\mathcal{O}(N_c^2N_t^2)$ of the multi-head self-attention (MSA) mechanism in transformer-based schemes. This indicates the proposed architecture is expected to become more advantageous in massive MIMO systems with more antennas and subcarriers. The measured average inference latency is only 15.88 ms. In addition, the overall complexity grows linearly with the number of users $K$ due to the weight-sharing strategy at the UEs and the multi-user detection mechanism at the BS.

\begin{table}[htbp]
	\centering
	\captionsetup{labelfont={color=red}}
	\caption{{Computational complexity comparison}}
	\setlength{\tabcolsep}{9pt} % 默认列间距设置为12pt
	\begin{tabular}{c|c|c|c}
		\hline
		\textbf{Scheme} & \textbf{Params} & \textbf{FLOPs} & \textbf{Inference Latency} \\
		\hline
		Proposed MAXIM & 10.54 M & 737.68 M & 15.88ms \\
		Transformer & 10.64 M & 202.00 M & 9.09 ms \\
		DNet & 14.41 M & 96.24 M & 6.23 ms \\
		PCA-HB & 0 & 25.84M & 7.21 ms \\
		\hline
	\end{tabular}
	\label{tab:complexity}
\end{table}

\section{Conclusion}
This paper proposes a task-oriented DJSCC framework for joint CSI feedback and hybrid beamforming in multi-user cmWave MIMO-OFDM systems. To address the heavy feedback overhead and severe path loss inherent in cmWave regimes, the framework integrates DJSCC encoding, DJSCC decoding and hybrid beamforming modules into a unified end-to-end neural network to maximize the downlink sum rate. This integrated design addresses the dual challenges of high-efficiency semantic feedback and cost-efficient beamforming in dense IoT networks. Specifically, the proposed framework employs the MAXIM architecture for both the encoder and the decoder. The DJSCC encoder network extracts essential channel semantics, which the DJSCC decoder network subsequently utilizes to directly design the hybrid beamforming matrices. This task-oriented design bypasses the intermediate CSI reconstruction stage, thus avoiding the associated performance degradation. Moreover, to exploit the cross-polarization correlations in dual-polarized massive MIMO-OFDM systems, the proposed CPI module employs a co-attention mechanism, which significantly enhances feature matching and mitigates compression-induced distortion. Simulation results demonstrate that the proposed scheme consistently outperforms existing DL-based baselines and conventional modular designs, especially in the case of low SNR and limited feedback overhead, confirming its effectiveness and robustness for future cmWave wireless communication systems.

\bibliography{reference}

\end{document}